\documentclass[pra,amsmath,twocolumn,showpacs]{revtex4-1}
\usepackage{bm}
\usepackage{amssymb}
\usepackage{graphicx}
\usepackage{xcolor}
\usepackage{amsmath,graphicx}
\begin{document}
\def\la{{\langle}}
\def\q{\quad}
\def\n{\\ \nonumber}
\def\ra{{\rangle}}
\def\Ep{{\mathcal{E}}}
\def\omga{{\epsilon}}
\def\t{{\tau}}
\def\rd{\color{black}}
\def\h{\hat{H}}
\title{Collective tunnelling of strongly interacting cold atoms in a double well potential}
%\author{D.Sokolovski}
\author {D.  Sokolovski$^{1,2}$, X. Gutie\'rrez de la Cal$^{2}$, and  M. Pons$^{3}$}
\affiliation{$^1$Departmento de Qu\'imica-F\'isica, Universidad del Pa\' is Vasco, UPV/EHU, Leioa, Spain}
\affiliation{$^2$ IKERBASQUE, Basque Foundation for Science, Maria Diaz de Haro 3, 48013, Bilbao, Spain}
%\author {M. Pons}
\affiliation{$^3$Departmento de F\' isica Aplicada I, Universidad del Pa\' is Vasco, UPV/EHU, Bilbao, Spain}
\begin{abstract}
It is known that under resonance conditions, a group of strongly interacting bosonic atoms, trapped in a double-well potential, 
mimics a single particle, performing Rabi oscillations between the wells.
By implication, all atoms need to tunnel at roughly the same time, even though 
the Bose-Hubbard Hamiltonian accounts only for one-atom-at-a-time transfers.
We analyse the mechanism of this collective behaviour,
evaluate the Rabi frequencies in the process, and discuss the limitation of this simple picture.
In particular, it is shown that the small rapid oscillations superimposed 
on the slow Rabi cycle, result from splitting the transferred cluster at the sudden onset 
of tunnelling, and disappear if tunnelling is turned on gradually.

\end{abstract}
%\pacs{PACS number(s): 03.65.Ta, 73.40.Gk}
\maketitle
%
%\large{
%
\section{Introduction}
For a quantum particle, which can occupy several quantum states with roughly the same energy, 
even a small perturbation is capable of causing transitions between the states.
If only a pair of states satisfies this resonance condition, these transitions would, in general, be in the form Rabi oscillations \cite{Rabi}, 
moving the particle periodically from one state to the other. 
A more interesting case is the one where the resonance occurs between many-body states of interacting particles, 
which correspond to different spatial configurations of the system.
A direct matrix element may or may not connect a pair of such states, say, $|\Psi_1\ra$ and $|\Psi_2\ra$. In the latter case the system
would need to reach the final state via a pathway, passing through the states of the system, whose energies can lie far from the resonance. The passage must be, in some sense, rapid, since a measurement will almost 
always find the system either in $|\Psi_1\ra$, or in $|\Psi_2\ra$, and only rarely detect it elsewhere.
% In this paper we will study the mechanism of such oscillations on, 
%evaluate the corresponding Rabi frequencies, and analyse the effect the intermediate states may have on the observed 
%Rabi probabilities. 
\newline
Examples of collective behaviour can be found, for example, in cold atom physics.
In \cite{Dud} the authors observed coherent many-body Rabi oscillations in 
electronics transitions of interacting rubidium atoms. More relevant to our analysis 
is the direct observation of 
 %(for a review see \cite{Legg}), where
correlated tunnelling of pairs of strongly 
interacting cold atoms,  reported in \cite{Bloch}.

Recently developed laser techniques are capable of trapping
such atoms in quasi one-dimensional traps \cite{Raiz}, and the dynamics of interacting atomic systems has been extensively studied both by solving the Schr\"odinger equation numerically (see, e.g. \cite{MB1}, \cite{MB2}, \cite{MB3}), and by using a simplified Bose-Hubbard model,
% \cite{BH1}, \cite{BH2}
as was done, for example, in {\rd \cite{Bloch}, \cite{BH1}, \cite{BH2}}. 
{\rd{In \cite{Ref2_5}, \cite{Ref2_6} and \cite{Ref2_7} experimental and theoretical analysis of  larger systems are presented, and the contributions from higher bands are included. In these systems, effects such as breathing are cradle modes can be observed as a consequence of this high-bands contributions, and tunnelling between wells as an effect of the lower modes. }}
{\rd{Similar dynamics may also be observed, for example, using mixtures of atoms, where different inter-species interaction regimes were studied
 %and higher band contributions were taken into account 
 \cite{Ref1_1}, 
 or spin chains \cite{Ref1_2}. Other theoretical studies, with a large number of particles,  address the dynamics of a BEC in a double-well using the Gross-Pitaevskii equations including many-body interactions studying the self-trapping effect induced \cite{Ref2_2} or different regimes, from coherent oscillations to their suppression when the number of bosons is high \cite{Ref2_3}. Important experimental results have also been obtained for BEC in double-well potentials, such as the first realization of a single Josephson junction \cite{Ref2_1} or the more recent \cite{Ref2_4} where the dynamical control of correlated tunnelling processes of strongly interacting particles is presented.
}}
\newline
The tunnelling frequencies for a symmetric trap were first evaluated in \cite{BH0}, where the authors relied on the time independent perturbation theory of \cite{Bern}, in order to obtain energy splitting between the resonant states.
In their follow up paper \cite{BH00}, the authors of \cite{BH0} studied time evolution of the average difference of the wells' populations,
$\delta n(t)$, for various initial conditions, and  analysed the frequency spectrum of quantum fluctuations, superimposed on the Rabi cycle. 
\newline 
However, this is not the whole story, and certain aspects of the collective tunnelling phenomenon require a further discussion.
In particular, in the Rabi oscillations one would  find all the transferred atoms in the same well, at all times. 
This suggests that the atoms must tunnel together, almost instantaneously, or at least during a time much shorter than the Rabi period  \cite{FOOTall}.
Neither the analysis of \cite{BH0}, \cite{BH00}, nor the form of the Bose-Hubbard Hamiltonian, which contains only single atom transfer terms, give an immediate clue as to how this may be possible. 
A study of the  mechanism of this rapid collective transfer, and identification of the relevant time scales is the first of our aims.
\newline
Furthermore, the picture in which a cluster of transferred atoms behaves as a single particle, performing Rabi oscillations, is only approximate,
and deserves further attention. The mean populations difference \cite{BH0}, \cite{BH00},  $\la \delta n(t) \ra= \sum_{n=0}^N[ p(n,t)-(N-n)(1-p(n,t)]$, where $p(n,t)$ are the probabilities for finding $n$ out of $N$ bosons in the right well at a time $t$, is a rather crude averaged quantity, and may not be best suited for such an analysis.
Quantum fluctuations, evident in the $\la \delta n(t) \ra$ \cite{BH00}, come from the directly measurable individual probabilities $p_n(t)$.
These, in turn, are absolute squares of the sums of the probability {\it amplitudes}, corresponding to elementary processes, such as transfer of a single atom  from one well to the other. Identification of interfering scenarios, responsible for the additional oscillatory patterns, specific to many body Rabi oscillations, is the second main subject of this paper. 
\newline
Fortunately, with the tunnelling matrix element small, and the Rabi period large, the required analysis can be carried out already in the first non-vanishing order of the time dependent perturbation theory. With its help, we will show that the largest contribution to the probability of the resonance transfer of $n$ atoms will come from a process, in which all $n$ bosons "jump"  together, 
almost instantly if compared to the Rabi period. 
We will also demonstrate that the additional oscillations  result from the processes, in which one or more atoms are split from the tunnelling cluster.  This breakup of the cluster can be related to a sudden onset of tunnelling at the start of the experiment. Our prediction that the oscillations will be quenched if tunnelling is turned on slowly, (compared to the characteristic "jump time"), can be subjected to experimental verification.
\newline
The rest of the paper is organised as follows. In Section II we formulate many-body resonance conditions  
for atoms in an asymmetric trap. In Sec. III we use time-dependent perturbation theory to analyse 
a rapid transfer of a group of atoms along an indirect pathway, connecting the resonance states, 
and obtain expressions for the Rabi frequencies.  In Sec. IV we relate additional oscillatory patterns, seen in the Rabi probabilities,
to sudden switching of the tunnelling. In Sec. V we briefly consider a detuned regime, and Section VI contains our conclusions.
A further detailed discussion of the interference mechanism of the transfer can be found in Appendix A. Our  findings are tested in the Appendix B on some of the exactly solvable cases.
%%%%%%%
\section {Trapped atoms in the Bose-Hubbard approximation}
%$\Omega /V<<1$}
We consider  $N$ strongly interacting identical bosons, contained  in an asymmetric double-well
potential, as shown in Fig.1. {\rd{An experimental realisation of such a system can be achieved, e.g.,  using $^{87}$Rb atoms (see, for example \cite{Ref2_1}).}}
The energy of the system is modelled by the Bose-Hubbard Hamiltonian  (we use $\hbar=1$)
%for the system is given by
\begin{eqnarray} \label{0.1}
\hat{H}_{}(t)= U(c_L^+)^2(c_L)^2/2+U(c_R^+)^2(c_R)^2/2+\n
\beta c_R^+c_R
+\Omega (c_L^+c_R+c_R^+c_L),\q\q\q
\end{eqnarray}
where $c_L^+$($c_L$) and $c_R^+$($c_R$) create (annihilate)
a boson in the left (right) state in Fig. 1.

{\rd{In the first two terms in Eq.(\ref{0.1}), we take  $U$ to describe short range interactions between the bosons in the same well,  which is neglected for atoms placed on different sides of the barrier}},
%The first two terms in Eq.(\ref{0.1}) describe the interaction between the bosons in the same well,
$\beta$ is the difference 
between one-particle energies on the right and on the left, and $\Omega$ is the tunnelling amplitude,
which allows for a transfer of a particle between the wells.
{\color{black}Note that both $U$ and $\beta$ can, in principle, be positive or negative, although in what follows 
$U,\beta >0$ will be considered.}
\newline
It is convenient to describe the system 
by the number of bosons populating the right well, $0\le n \le N$.
If no tunnelling is possible, $\Omega=0$, the eigenstates of $\hat{H}_{}$
%are 
\begin{eqnarray} \label{0.1a}
|n\ra = (c_L^+)^{N-n}(c_R^+)^n|0\ra/[n!(N-n)!]^{1/2}
\end{eqnarray}
correspond to the energies 
\begin{eqnarray} \label{0.1b}
%\Ep(n,\nu)=Un(n-\nu)+\Ep_0, \q n=0,1,...,N\n
\Ep(n)=U(n-\nu)^2+\Ep_0(N,\beta), \q n=0,1,...,N,
\end{eqnarray}
where
\begin{eqnarray} \label{0.1d}
\nu \equiv (N-\beta/U)/2, \q \text{and} \q \Ep_0\equiv U[N(N-1)/2-\nu^2].
\end{eqnarray}
%shown in Fig.2. 
The dependence of $\Ep(n)$ on $n$ is quadratic and, since $\Ep(n)=\Ep(2\nu-n)$, 
whenever the minimum of the parabola, at $n=\nu$ is an integer, 
or half integer, 
\begin{eqnarray} \label{0.1c}
%\Ep(n,\nu)=Un(n-\nu)+\Ep_0, \q n=0,1,...,N\n
 (N-\beta/U)=K, \q K=1,2,...
\end{eqnarray}
there are several pairs of doubly degenerate states. 
{\rd{Fig. 2 shows the energy levels for an asymmetric potential with $N=8$ and $\beta/U=1$, so the minimum energy corresponds to $\nu=3.5$. We have degeneracy between several pairs of states, for instance, between $|1\ra$ and $|6\ra$, that share the same energy, $\Ep(1)=\Ep(6)$.
 In general, for $\nu \le N/2$ the number of degenerate pairs
is given by  that of the integers in the interval $[0,\nu)$, and 
for $\nu > N/2$, by the number of integers inside $(\nu/2,N]$.}}
 We note that if $\nu$ happens to be an integer, there is  an unpaired  non-degenerate ground state
$|n=\nu\ra$, and $\Ep( \nu)= \Ep_0$. 
\newline
Next we switch the tunnelling on, in such a manner that the tunnelling matrix element in Eq.(\ref{0.1}) will remain small, 
compared to the interatomic interaction
$\Omega /U <<1$.
The perturbation will lift the degeneracy
between the levels 
$|n\ra$ and $|2\nu-n\ra$, and introduce pairs of new eigenstates 
\begin{eqnarray}\label{0.2}
|n_\pm\ra=[|n\ra \pm |2\nu-n\ra]/\sqrt 2
\end{eqnarray}
with the energies
\begin{eqnarray}\label{0.3}
\Ep(n_\pm)\equiv \Ep(n)\pm \omega_n = \Ep(2\nu -n)\pm \omega_n,
\end{eqnarray}
with the  splitting $2\omega_n$ small, compared to other energy
differences shown in Fig. 2.
\newline
%As a result, the atoms, initially prepared in the
%state $|\Psi(0)\ra=|n\ra$ with $n$ atoms in the right well, will perform collective Rabi oscillations
%between the states $|n\ra$ and $|2\nu-n\ra$, so that their state at a time $t$ will be given by
{\color{black} If as a result, the atoms, initially prepared in the
state $|\Psi(0)\ra=|n\ra$ with $n$ atoms in the right well, will perform collective Rabi oscillations
between the states $|n\ra$ and $|2\nu-n\ra$, we expect their state at a time $t$ to be given by}
\begin{eqnarray}\label{RAB}
|\Psi(t)\ra\approx [\cos(\omega_nt)|n\ra+\q\q\q\q\q\q\n
\alpha \sin(\omega_nt)|2\nu-n\ra]\exp(-i\Ep(n) t), \quad  |\alpha|=1.
\end{eqnarray}
For a symmetric setup, the frequencies $\omega_n$ were obtained in \cite{BH0} within time-independent 
perturbation theory, using a rather complicated procedure \cite{Bern} for diagonalising a 
tridiagonal matrix with degenerate eigenvalues. 
The authors of \cite{BH0} correctly note that the shortest way to reach the state $|N-n\ra$ from $|n\ra$
is by "moving one boson per step". There is, however, room for a further clarification.
In the approximation (\ref{RAB}), the atoms will never be observed
 in any other state
$|m\ra$, $m\ne n,2\nu-n$  (or rather the probability of such an observation
will be negligibly small).
For example, in the case shown in Fig. 2, for $|\Psi(0)\ra=|1\ra$, counting atoms
in the right well will  almost always give a result either $1$, or $6$. This suggests,
and we will show below that this suggestion is correct, that the $5$ atoms 
making the difference, will need to tunnel, in some sense,  {\it  all at the same time}.
Next we  demonstrate this by using time-dependent perturbation theory,
and obtain, while we are at it, collective tunnelling frequencies for an asymmetric trap ($\beta \ne 0$).  

%%%%%%%%%44$$$$
\begin{figure}[h]
\includegraphics[angle=0,width=7cm, height= 5cm]{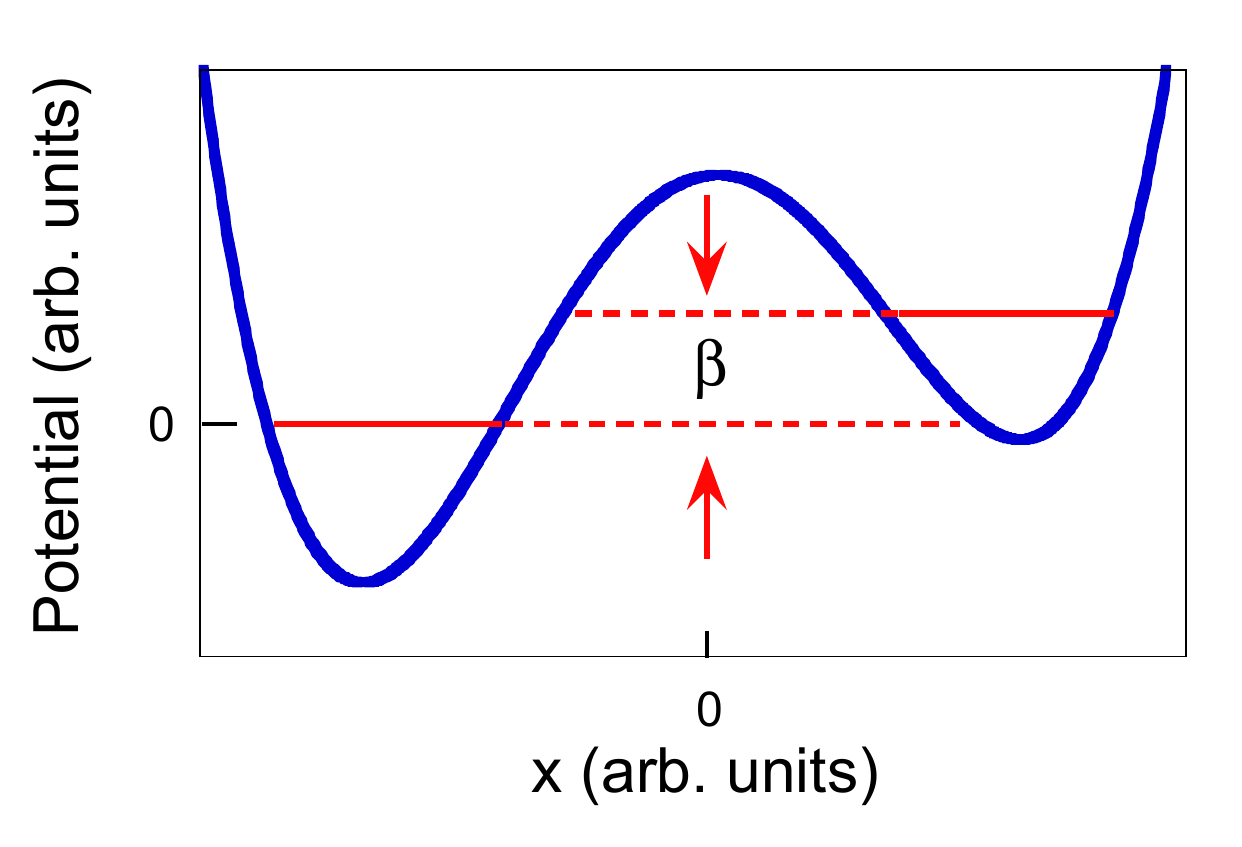}
\caption {An asymmetric double-well potential supports two single-particle levels, 
with an energy difference $\beta$. Tunnelling across the barrier 
is described by the amplitude $\Omega$ in Eq.(\ref{0.1}). }
\label{fig:FIG1}
\end{figure}
\begin{figure}[h]
%\vskip0.5cm
\includegraphics[angle=0,width=7cm, height= 8cm]{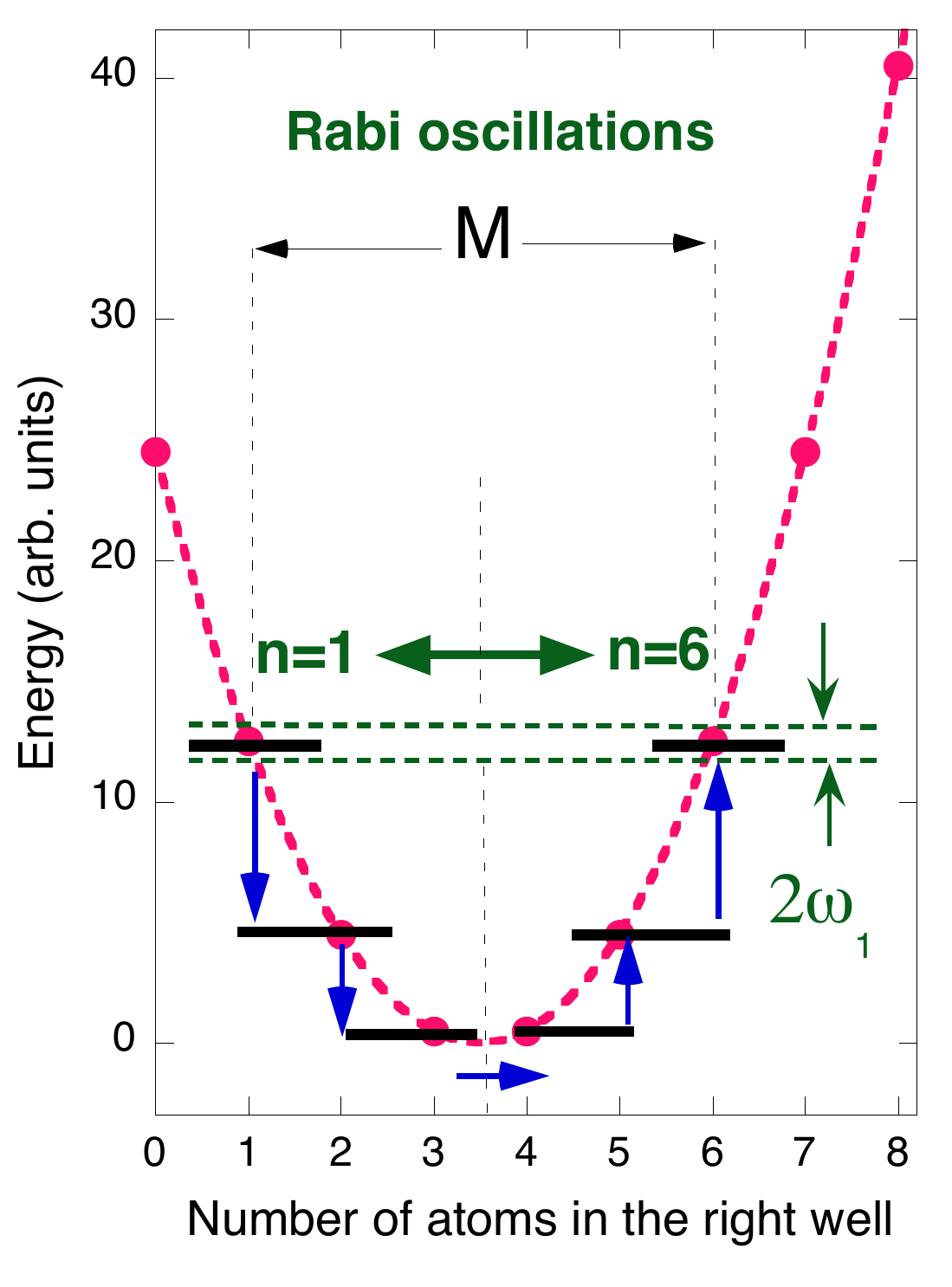}
\caption {Energy levels for strongly interacting bosonic atoms, with $N=8$ and {\rd{$\beta/U=1$}}, trapped in an asymmetric potential well. 
Degeneracy between the states with $n=1$ and $n=6$ atoms in the right well is lifted by tunnelling,
which causes periodic transfer of $M=5$ atoms between the wells. No matrix 
element connects the levels directly, and the transition has to follow the pathway shown by the arrows.
In the strong coupling limit, all five atoms tend to tunnel together.}
% although the moment at which the transition takes place remains 
%indeterminate. }
\label{fig:FIG1}
\end{figure}
%%%%%%%%%%%%%%%%%%%%%%%%%%%%%
\section{Perturbation theory and the collective frequencies}
Writing the wave function of the system as
{\color{black}
\begin{eqnarray} \label{1.1}
|\Psi(t)\ra = \sum_{n=0}^N b_n(t) |n\ra,
\end{eqnarray}
%with an initial condiion $b_m(0)=\delta_{nm}$,}
%\vskip0.5cm
and introducing dimensionless time and energy,
$$ \tau \equiv Ut, \q E_n\equiv \Ep(n) /U ,$$
%and measuring energy from $N(N-1)U$,
we obtain the equations for the coefficients  $b_n(\tau)$ 
\begin{eqnarray} \label{1.2}
i\partial_{\tau}b_n=\sum_{m=0}^N(E_n\delta_{nm}+\gamma_{n-1}
\delta_{n,m+1}+\gamma_{n}
\delta_{n,m-1})b_m\q
\end{eqnarray}
where
%$$\gamma_n\equiv \Omega/U [n(N-n)]^{1/2} ????$$
\begin{eqnarray} \label{1.2a}
\gamma_n\equiv U^{-1}\Omega [(n+1)(N-n)]^{1/2}.
\end{eqnarray}
{\color {black}A probability to find $n$ atoms in the right well is, therefore, given by
\begin{eqnarray} \label{1.1s}
p(n)=|b_n(t)|^2.
\end{eqnarray}}
Equations (\ref{1.2}) can be solved analytically only for $N\le 3$ in general, and for $N\le 7$ if the potential is symmetric, as discussed in Appendix B. In order to evaluate the Rabi frequencies for an arbitrary $N$,
we note that since no matrix element connects the states $|n\ra$ and $|2\nu-n\ra$ directly,
and the latter can only be reached from the former via $2(\nu-n)$ intermediate 
steps, shown in Fig. 2. 
%Thus we expect the energy splitting and the frequency 
%$\omega_n$ to be of order $\gamma^{2(\nu-n)}$, and will employ the time-dependent perturbation theory, in order to provide also  an
%additional insight into the simultaneous nature of collective tunnelling.
\newline
{\color {black}For simplicity, we first consider the symmetric potential, $\beta=0$, 
assume, for the moment,
% that $\Omega$ depends on time, so that 
%the time dependence of 
that the $\gamma$s depend on time, let the system start with $n$ bosons in the right well,
and employ the time-dependent perturbation theory.}
To the leading approximation, the transition amplitude between the states $|n\ra$ and $|N-n\ra$
%after a time $\tau$ 
 is given by ($M\equiv N-2n$, $n\le N/2$)
\begin{eqnarray} \label{0.4}
A_{n,N-n}(\tau)\equiv\la N-n|\exp[-i\int_0^\tau H(\tau')d\tau']|n\ra =\n
(-i)^M\int_0^\tau d\t_M...\int_0^{\t_3}d\t_2 \int_0^{\t_2}d\t_1\times\q\q\q\n
 \exp[-iE_{N-n}(\t-\t_{M})]\gamma_{N-n-1}(\t_M)\q\q\q\n
\times \exp[-iE_{N-n-1}(\t_{M}-\t_{M-1})]...\q\q\q\q\n
%\nonumber
\times \gamma_{n+1}(\t_2)\exp[-iE_{n+1}(\t_2-\t_1)]\q\q\q\q\n
\times \gamma_n(\t_1)\exp(-iE_n\t_1)+O(\Omega^{M+1}/U^{M+1}).
\end{eqnarray}
%where $\t\equiv TU$.
{\color {black}This expression has the standard interpretation \cite{Feyn}:}
the first of the $N-2n$ atoms jumps into the right well at $t_1$,
the second at $t_2$, and so on. The transition amplitude is then found
by summing over all $t_j$, leaving the precise moments of jumps
indeterminate. {\color {black} Perturbative treatment will limit us to 
times much shorter that the Rabi period, yet as we will see below, 
it captures the essential features of the collective transfer mechanism.}
\newline
Next we demonstrate that for $\Omega/U << 1$, {\color {black} and $\Omega(t)$ slowly varying 
compared to the inverse of the separations between the system's levels in Fig.2,} the sum is dominated 
by the process in which all $N-2n$ atoms jump roughly at the same time.
Returning to the original unscaled time variable, $t$, 
we have
\begin{eqnarray} \label{0.5}
A_{n,N-n}(T) =
(-i)^M\exp[-i\Ep(N-n)T]\q\q\q\q\n 
\times\int_0^Tdt_M W_{N-n-1}(t_M)\exp[i\Delta_Mt_M]...\q\q\q\q\\
\nonumber
\times\int_0^{t_3}dt_2W_{n+1}(t_2)\exp[i\Delta_2 t_2]\q\q\q\q\q\q\q\q\n
\times\int_0^{t_2}dt_1W_n(t_1)\exp(i\Delta_1 t_1)+O(\Omega^{M+1}/U^{M+1})
\end{eqnarray}
{\color {black} where $\Delta_m$ is the energy, separating two adjacent states,} 
\begin{eqnarray} \label{0.5a}
%\nonumber
\Delta_m\equiv \Ep(n+m)-\Ep(n+m-1)\n
=U[1-2(n+m)-N], 
%W_n(t)=\gamma_nU= \Omega(t) [(n+1)(N-n)]^{1/2}.\q\q\q
\end{eqnarray}
{\color {black}and $W_n(t)\equiv \gamma_nU= \Omega(t) [(n+1)(N-n)]^{1/2}$ is the scaled 
tunnelling amplitude.}
%$$\Ep_n=E_nU=-Un(N-n)$$
%$$ \Delta_m\equiv \Ep_{n+m}-\Ep_{n+m-1}, \quad m=1,2,..M.$$
%$$W_n(t)=\gamma_nU= \Omega(t) [n(N-n)]^{1/2}$$
Thus, we have to evaluate $M$ oscillatory integrals. 
Oscillations of $\exp(i\Delta_n t)$ become more rapid as $U$ increases,
so that, for a large $U/\Omega$,  the main contributions to the integrals will come from the
endpoints (for details see Appendix A). For example, we may write
\begin{eqnarray} \label{int}
\int_0^{t_2}dt_1W_n(t_1)\exp(i\Delta_1 t_1)= \q\q\q\n
[W_n(t_2)\exp(i\Delta_1 t_2)-W_n(0)]/i\Delta_1+O(\Omega/U).
\end{eqnarray}
Here the first term corresponds to the first particle jumping at the 
same time as the second, $t_1=t_2$. The second term clearly corresponds
to the first jump occurring at $t_1=0$, i.e., immediately after the tunnelling
is switched on.
Jumps at $0 < t_1<t_2$  are suppressed,  for large $U/\Omega$, due to destructive 
interference.
Continuing  in the same vein, we obtain a total of $2^M$ terms ranging 
from all particles jumping at $t=0$ to all particles jumping at the same time.
In the case of an exact resonance,
$\sum_{m=1}^M \Delta_m=0$,
 the contribution from all fully coordinated jumps between $t=0$ and $t=T$ is readily seen to be
\begin{eqnarray} \label{0.6}
i(-1)^M\int_0^Tdt
\prod_{j=0}^{M-1}W_{n+j}(t)\times\q\q\q\q\n
[\Delta_1(\Delta_1+\Delta_2)....
(\Delta_1+\Delta_2+...+ \Delta_{M-1})]^{-1}.\n
\end{eqnarray}
Note that since the amplitude results from the interference between all $t$'s, 
the exact moment in which the collective transfer takes place remains 
indeterminate, much like the number of the slit chosen by an electron 
in Young's double-slit experiment.
Returning to the case of constant $\Omega$, we, therefore, obtain
\begin{eqnarray} \label{0.7}
A_{n,N-n}(T) =i(-1)^MT
\frac{\Omega^M}{U^{M-1}}\exp[-i\Ep(N-n)T]\times \q\q\n
 \frac{(n+1)...(N-n)}{(E_{n+1}-E_n)
%(E_{n+2}-E_{n})
....(E_{N-n-1}-E_n)} +R,\q\q\q
\end{eqnarray}
where the remainder $R$
 contains the terms corresponding to, at least, one atom jumping 
 immediately after tunnelling is turned on.
 In the next Section we will demonstrate that  $R$ will vanish if 
 tunnelling is switched on sufficiently slowly. 
 % $\gamma(t=0)=\Omega(t=0)=0$ (see Appendix B).
\newline
For a sufficiently small $T$, Eq.(\ref{RAB}) predicts
\begin{eqnarray} \label{0.8}
A_{n,N-n}(T) =\alpha \omega_n T.
\end{eqnarray}
Comparing Eq.(\ref{0.8}) with Eq.(\ref{0.7}), and evaluating the products, yields
%\begin{eqnarray} \label{0.9}
%\omega_n=\frac{\Omega^{N-2n}}{U^{N-2n-1}}
%\frac{(n+1)(n+2)...(N-n)}{(E_{n+1}-E_n)(E_{n+2}-E_{n})....(E_{N-n-1}-E_n)},
%\quad \alpha=(-1)^{N-2n}i.
%\end{eqnarray}
%or, equivalently,
\begin{eqnarray} \label{0.10}
\omega_n=\frac{\Omega^{N-2n}}{U^{N-2n-1}}
\frac{(N-n)!}{n![(N-2n-1)!]^2},
\quad \alpha=(-1)^ni,\q
\end{eqnarray}
which agrees with the result obtained by a different method in \cite{BH0}.
A calculation  for an asymmetric well can be done in exactly the same way, and here we will 
only quote the final result. As discussed in Sec. II, the resonances between
many-body states occur provided $\nu=1/2,1, ...,N/2$. The Rabi 
frequency for a process in which the atoms start in a state $|n\ra$, and $2(\nu-n)$,
[$0\le n\le \nu-1$ if $\nu$ is an integer, and $0\le n\le \nu-1/2$, if $\nu$ is odd]
where the atoms are transferred simultaneously, is given by [$\nu\equiv (N-\beta/U)/2$]
\begin{eqnarray} \label{0.10a}
\omega_n=\frac{\Omega^{2(\nu-n)}}{U^{2(\nu-n-1/2)}[2(\nu-n)-1]!^2}
%\times\n
\sqrt{\frac{(N-n)!(2\nu-n)!}{n!(N+n-2\nu)!}}.\q\q\n
%\q\q\q\q\q\q\q\q.
\end{eqnarray}
For an asymmetric well, the leading probabilities $p(n)$, to find $n$ atoms in the right well, 
after switching the tunnelling on at $t=0$,
are shown 
in Fig.3, together with the Rabi oscillations at the frequency (\ref{0.10a}).
% assuming that tunnelling is switched on suddenly at $t=0$.
Superimposed upon the Rabi oscillations, there are much faster oscillations (see inset), which will be discussed in the next Section.
\begin{figure}[h]
\includegraphics[angle=0,width=8.5cm, height= 7cm]{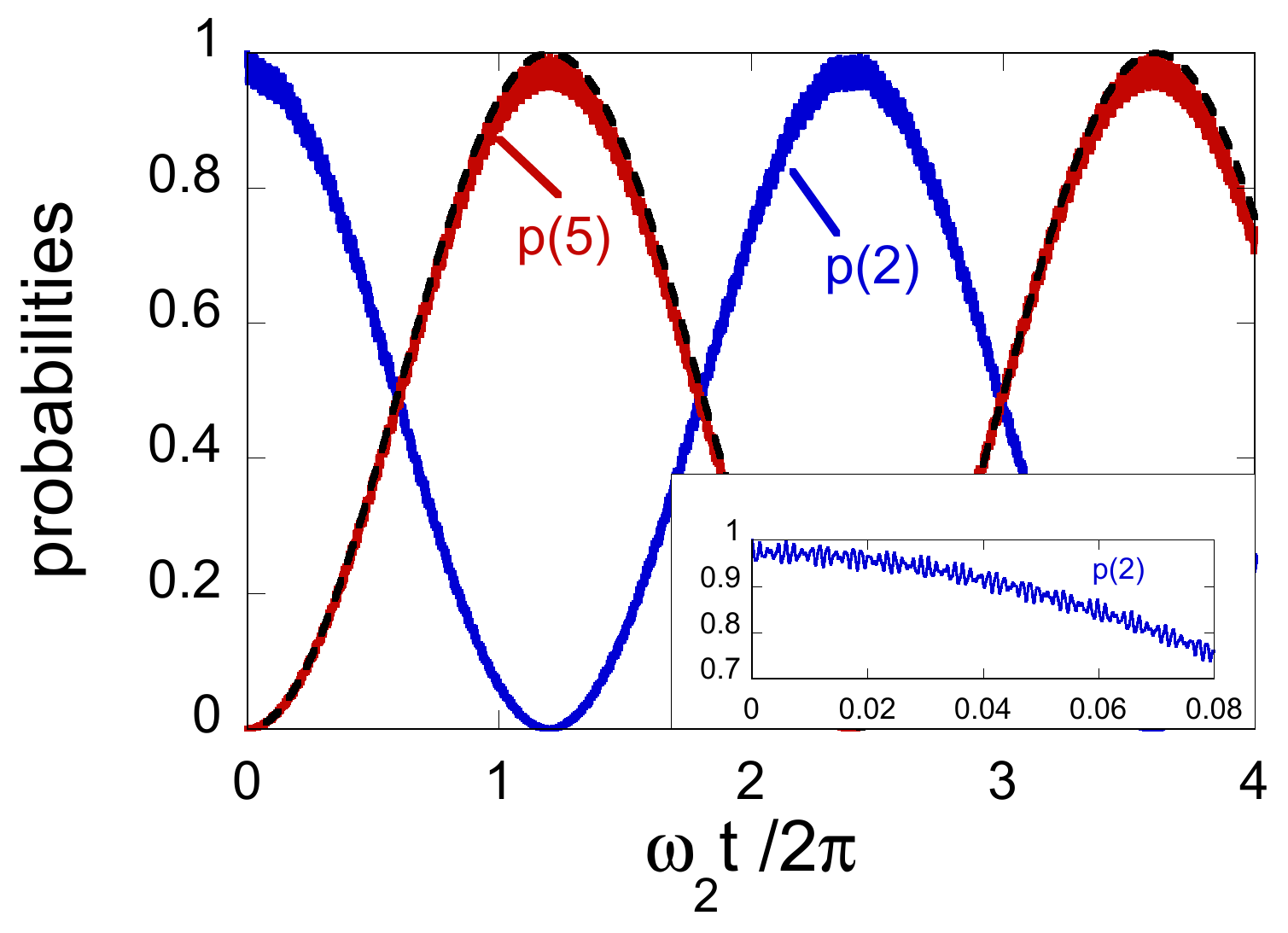}
\caption {The leading probabilities $p(2)$ and $p(5)$, for collective transfer of $M=3$ out of $N=8$
atoms, and  
 $\beta/U=1$ {\rd{(c.f. Fig.2)}}, and $\Omega/U=0.1$, {\color {black} obtained by numerical integration of Eqs.(\ref{1.2})},
Also shown by a dashed line is the Rabi approximation $p(2)=\sin^2(\omega_2 t)$, with $\omega_2$ given 
by Eq.(\ref{0.10a}).
Remaining  probabilities $p(m)$, $m\ne 2,5$ are too small to be shown on the same scale.
The inset gives a more detailed view of the rapid oscillations superimposed on the Rabi probabilities. }
\label{fig:FIG1}
\end{figure}
%%%%%%%%%%%%%%%%%%%%%
\begin{figure}[h]
\includegraphics[angle=0,width=9cm, height= 8cm]{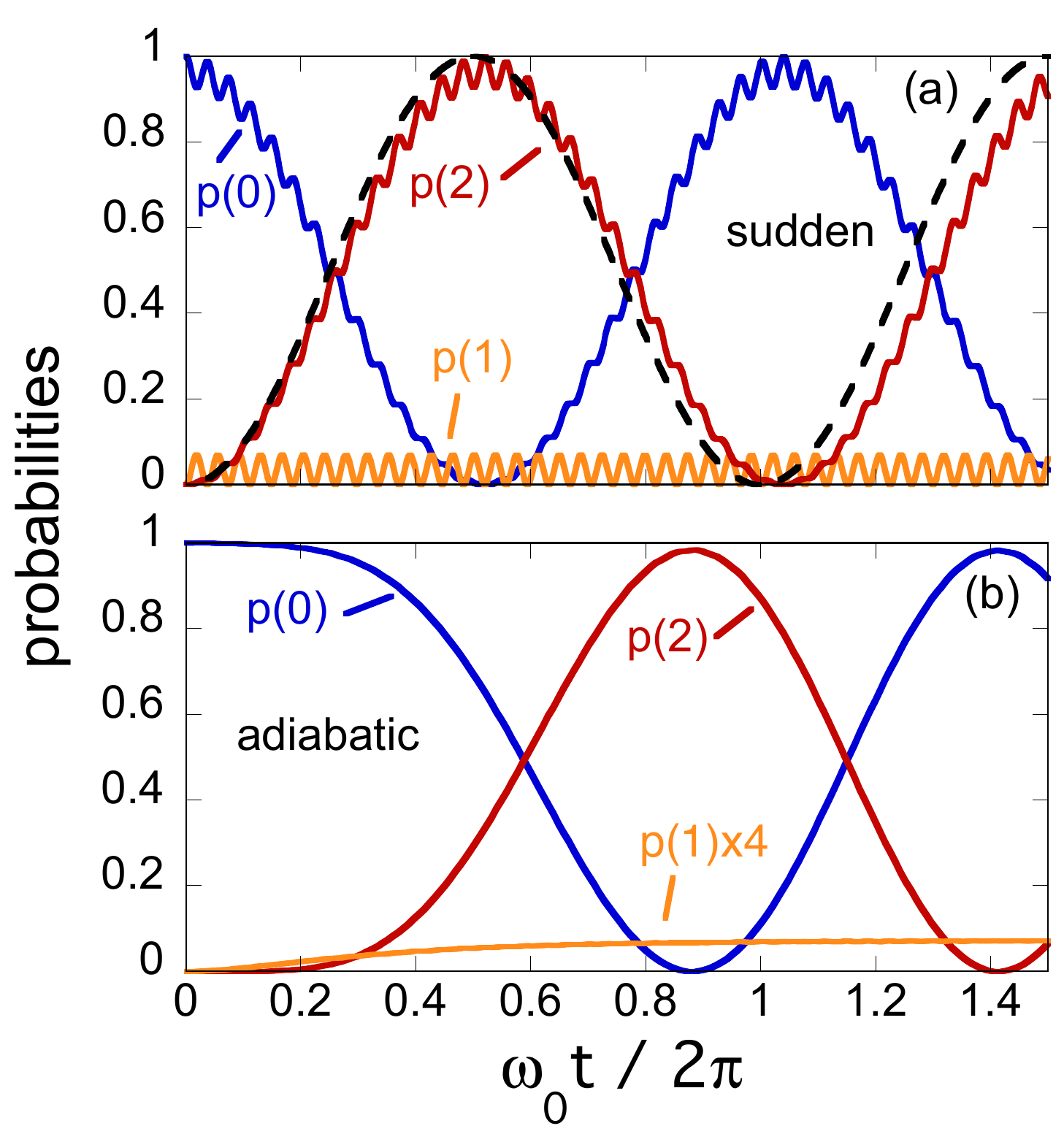}
\caption {a) All three probabilities,  $p(0)$, $p(1)$ and $p(2)$, for a 
coherent transfer of $M=2$ out of $N=2$ atoms,
and  $\Omega/U=0.1$, $\beta/U=0$, 
with the tunnelling switched on suddenly at $t=0$.
Also shown by a dashed line is the Rabi approximation $p(0)=\sin^2(\omega_0 t)$,
with $\omega_0$ given 
by Eq.(\ref{0.10}).
b) Same as a) but with the tunnelling switched on gradually,
with $\Gamma$ in Eq.(\ref{int1aa}) chosen so that $\gamma /U=0.1$}
\label{fig:FIG1}
\end{figure}
%%%%%%%%%%%%%%%%%%%
\begin{figure}[h]
\includegraphics[angle=0,width=8cm, height= 6cm]{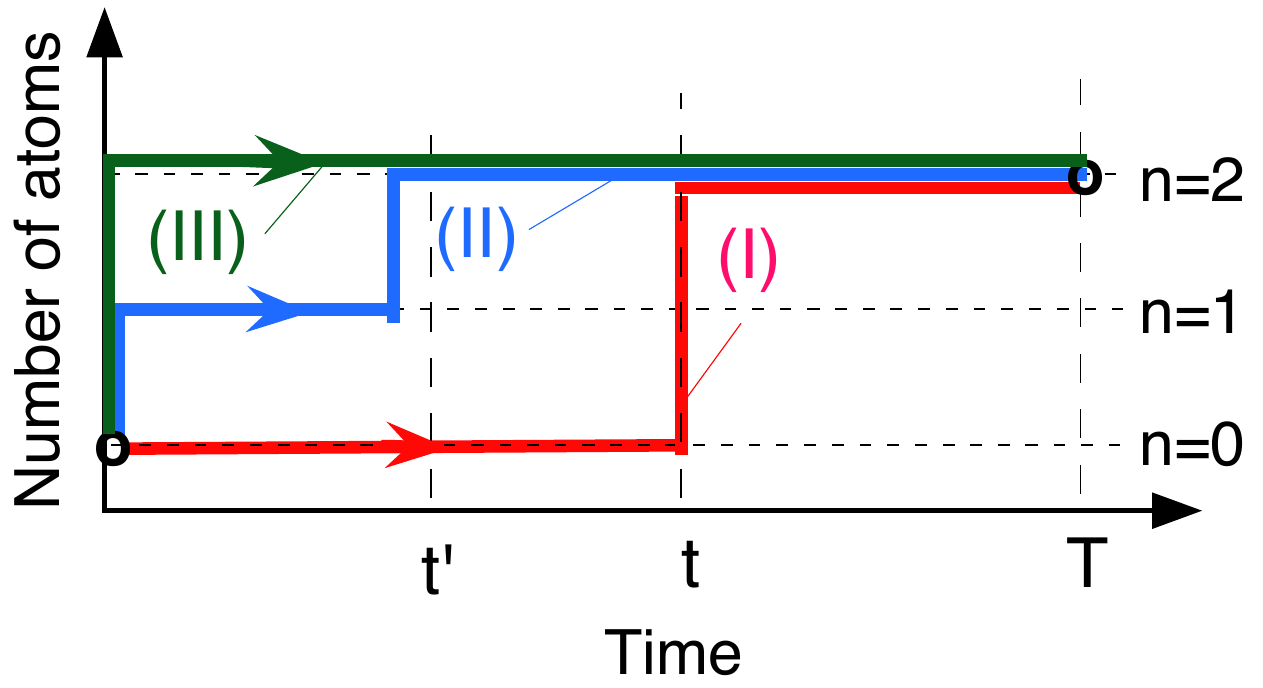}
\caption {Three ways to transfer $M=2$ out of $N=2$ atoms from the left to the right well.
(I) both  atoms are transferred at the same time $t>0$; (II) the first atom jumps at
$t=0$, and the remaining one is transferred together at some $t'$,
(III) all atoms are transferred at $t=0$, when the tunnelling is switched on, 
The second and third processes are quenched if the tunnelling matrix element 
$\Omega(t)$ is switched on slowly from zero.}
\label{fig:FIG1}
\end{figure}
%%%%%%%%%%%%%%%%%%%%%%%%%%%%%%%%%
\section{Sudden vs. adiabatic switching on of the tunnelling}
These small rapid  oscillations are much more pronounced in a system containing just $N=2$ atoms, as shown in Fig. 4a.
%The system with no atoms in the right well, $n=0$ experiences a sudden onset of tunnelling, resulting mainly in a coordinated transfer of two atoms, while the probability to find only one atom in the right well is small. 
We have already used perturbation theory to obtain slow Rabi frequencies $\omega_n$, and next we will
try to use it in order to explore the origin of these additional oscillatory patterns. 
\newline
To the leading order in $\Omega$, the amplitude to find one atom in each well is given 
by Eq.(\ref{B1}),  and we have (from (\ref{0.1b}), $\Ep(0)=U$, and $\Ep(1)=0$)
 \begin{eqnarray} \label{2.2}
A_{0,1}(T)  \approx \n
-i \sqrt 2 \Omega\int_0^T dt 
\exp[-i\Ep(1)(T-t)]\exp[-i\Ep(0)t] \n
= \frac{2\sqrt 2 \Omega}{U}\exp(-iUT/2)\sin(iUT/2), 
\end{eqnarray}
so that the corresponding probability rapidly oscillates,
 \begin{eqnarray} \label{2.2a}
p(1)  \approx 
{8 \Omega^2}/{U^2}\sin^2(iUT/2). 
\end{eqnarray}
As discussed in Appendix A, by turning the tunnelling on gradually, we can avoid splitting the $2$-atom cluster at 
$t=0$, and quench the oscillations. This can be checked directly, e.g., by choosing 
\begin{eqnarray} \label{int1aa}
\Omega(t)=\Omega[1-\exp(-\Gamma t)], 
\end{eqnarray}
{\rd{where $1/\Gamma$ is the characteristic switching time, over which the tunnelling matrix
element is brought to its stationary value.}} Evaluation of the integral in Eq.(\ref{2.2}) gives
\begin{eqnarray} \label{int1}
A_{0,1}(T) \approx 
%\int_0^{t_2}dt_1W_n(t_1)\exp(i\Delta_1 t_1)=\q\q\q\q\q\q\q\q\\
%\nonumber
\sqrt 2 \Omega  \{\exp(-iU T)/U-\q\q\q\q\q\n
\exp[-i(U-i\Gamma) T]/(U-i\Gamma)
-1/U
+1/(U-i\Gamma)\}.\q\q
\end{eqnarray}
For $\Gamma /U<<1$, and $\Gamma T>>1$, this reduces to
\begin{eqnarray} \label{int2}
A_{0,1}(T) \approx \sqrt 2 \Omega/U  \{\exp(-iU T), \q\text {and}\q p(1) \approx 2\Omega^2/U^2.\q\q
\end{eqnarray}
Oscillations of $p(1)$ disappear, and its largest value is reduced by a factor of four.
The rapid oscillations also disappear from  the amplitude to find $2$ atoms in the right well.
This, as described in Appendix A, can be written as a sum of three terms,
\begin{eqnarray} \label{2.3}
A_{0,2}(T)  
\approx A^{(I)}_{0,2}(T) +A^{(II)}_{0,2}(T)+A^{(III)}_{0,2}(T)=\n
i\frac{2\Omega^2}{U}\exp(-iUT)T
-\frac{2\Omega^2}{U^2}
+\frac{2\Omega^2}{U^2}\exp(-iUT), 
\end{eqnarray} 
related to the three scenarios shown in Fig.5. The first term corresponds to both atoms being transferred
together at some unspecified time between $t=0$ and $t=T$. The second term comes from the process 
in which one of the atoms jumps immediately, at $t=0$, while the second one is transferred later.
The third term is the amplitude for both atoms jumping together at $t=0$
(see Appendix A). With tunnelling switched on gradually from zero we, therefore, have 
\begin{eqnarray} \label{int3}
A_{0,1}(T) \approx A^{(I)}_{0,2}(T) ,\q\text {and}\q p(1) \approx 4\Omega^4T^2/U^2.
\end{eqnarray}
\newline
What has been said so far should apply to times much shorter than the large Rabi period,
$T_{Rabi}=2\pi/\omega_0$, $ 1/U<< T<< T_{Rabi}$, and we still need to check that
this analysis remains correct for much longer times, shown in Fig. 4. We note first that the exact 
amplitude [cf. Eq.(\ref{A1b})] has a form, similar to $A_{0,2}$ in Eq.(\ref{2.2})
 \begin{eqnarray} \label{2.4}
A_{0,1}(T)  = 
 -\frac{2\sqrt 2 \Omega}{\kappa U}\exp(-iUT/2)\sin(i\kappa UT/2), 
\end{eqnarray}
where  $\kappa =\sqrt{1+16\Omega^2/U^2}$. The amplitude remains small, $\sim \Omega$, 
at all times, since the system never stays in the one-atom-in-each-well configuration for long.
We can, therefore, expect  that, in the slow switching-on regime, the probability $p(1)$ would reduce to 
 \begin{eqnarray} \label{2.5}
p(1) = \frac {2\Omega^2}{\sqrt{U^2+16\Omega^2}},
\end{eqnarray}
and the oscillations will be eliminated also from $p(0)$ and $p(2)$, 
which depend on $p(1)$, through the condition $p(0)+p(1)+p(2)=1$.
The ultimate proof consists in solving numerically Eqs.(\ref{A1a}) with an $\Omega(t)$ given in Eq.(\ref{int1aa}).
The results are shown in Fig.4b. The Rabi oscillations, delayed until  $\Omega$ reaches its final magnitude, 
lack the rapid oscillations seen in Fig. 4a, while $p(1)$ does indeed have a constant value given by (\ref{2.5}).
This approach can be extended, with similar results, to more than two atoms.
Before giving our conclusions, we briefly discuss detuned Rabi oscillations.
%%%%%%%%%%%%%%%%%%%%%%
\section{Detuned Rabi oscillations} 
At the exact resonance, $\Ep(n)=\Ep(2\nu -n)$
Rabi oscillations between the corresponding states 
$|n\ra$ and $|2\nu-n\ra$, $\nu=(N-\beta/U)/2$,
can be described by an effective Hamiltonian, {\color {black}  acting in a reduced two-dimensional Hilbert 
space, 
$\h^{eff}=\omega_n\sigma_x$, with $\sigma_x$ denoting the Pauli matrix. } 
In a slightly asymmetric well, with $\beta$ replaced by $\beta +\delta \beta$,
%$\beta << U$, 
the two levels are detuned by $\Delta \Ep$, $|\Delta \Ep|=|2\delta \beta (n-\nu)|$
%$2(\nu-n)\beta=2\beta$, 
and the effective Hamiltonian acquires an additional term, proportional to $\sigma_z$, 
$\h^{eff}=\omega_n\sigma_x+\Delta \Ep \sigma_z/2$. Here, as in the general case, the 
probabilities are given by the formulae, known for detuned Rabi oscillations. 
In particular, starting with $n$ atoms in the right well we should have
\begin{eqnarray} \label{2.2aa}
P(2\nu-n)=1-P(n)\approx \q\q\q\n
\frac{\omega_n^2}{\omega_n^2+\delta \beta^2 (\nu-n)^2}
\sin^2\left[\sqrt{\omega_n^2+{\delta \beta^2 (\nu-n)^2}t}\right],\n
P(m)<< 1, \q \text{for}\q m\ne n, 2\nu-n\q\q
 \end{eqnarray}
where $\nu$ and $\omega_n$ are given by Eqs.(\ref{0.1b}) and (\ref{0.10a}), respectively.
Detuned oscillations for $N=5$, $n=0$, in a slightly perturbed symmetric potential
%$\beta=...$, and $\delta \beta = ...$
%%%%%%%%%%%%%%%%%%%
\begin{figure}[h]
\includegraphics[angle=0,width=8cm, height= 6cm]{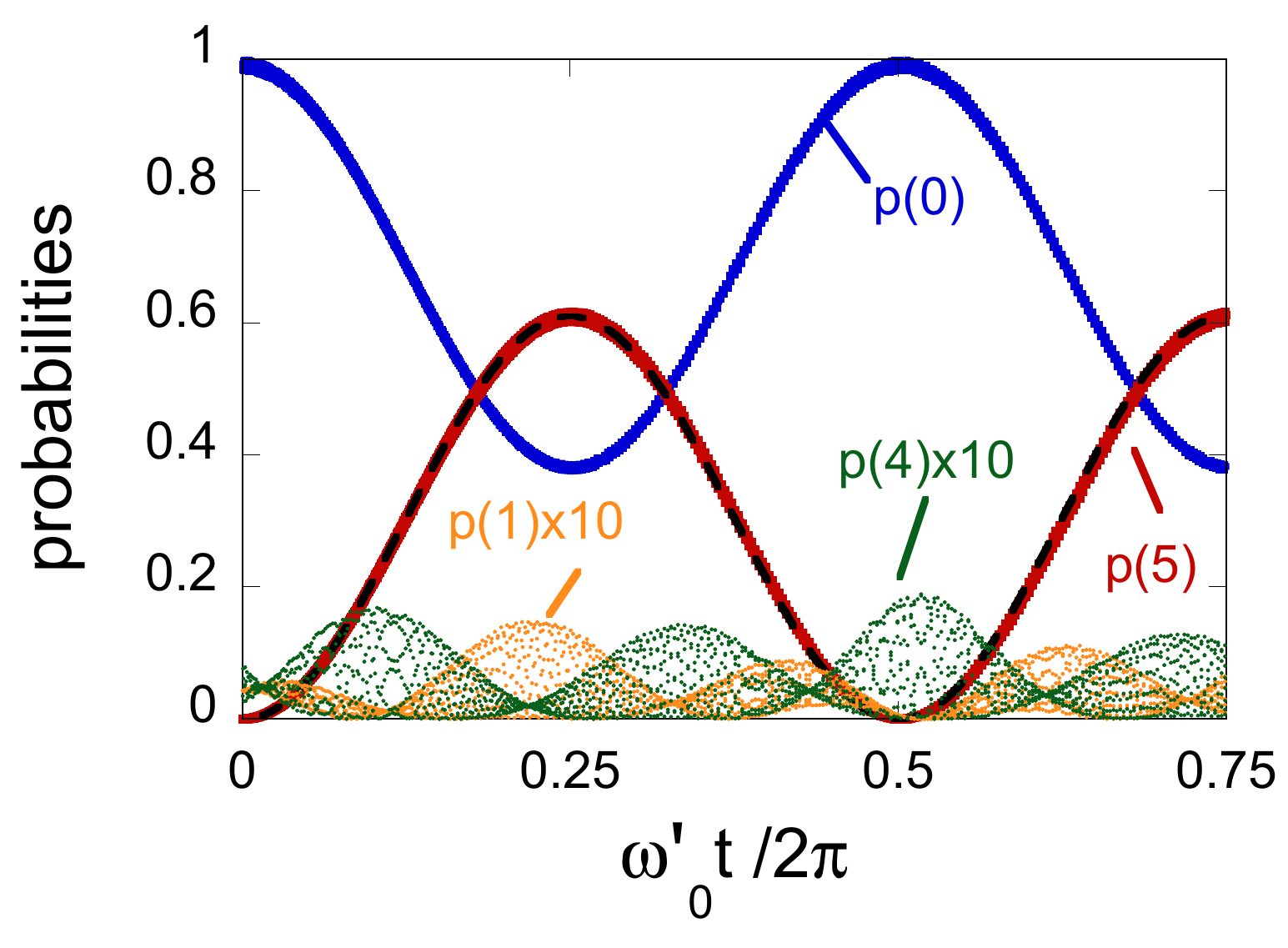}
\caption {Four leading probabilities, {\color {black} obtained by numerical integration of Eqs.(\ref{1.2})} [$p(1)$ an $p(4)$, (dots) are magnified for better viewing],
for coherent transfer of all $M=N=5$ atoms 
in a slightly asymmetric double-well, $\beta/U=0$, $\delta \beta/U=4*10^{-6}$, 
and $\Omega/U=0.1$. The frequency of the detuned Rabi oscillations {\color {black} (dashed)}, 
$\omega'=\sqrt{\omega_n^2+{\delta \beta^2 (\nu-n)^2}}$ is very close to $\omega_0$.}
\label{fig:FIG1}
\end{figure}
%%%%%%%%%%%%%%%%%%%%%%%%%%%%%%%%%
are shown in Fig. 6.
\section{Conclusions and discussion}
In summary, we show that a system of strongly interacting bosonic atoms in a not necessarily symmetric double-well potential is capable 
of performing collective Rabi oscillations of a frequency $\omega_n$ given by Eq.(\ref{0.10a}).
This will happen provided the resonance condition (\ref{0.1c}) is satisfied for the many body states containing $n$ and $n+M$ atoms in one of the wells. 
The frequencies of the Rabi oscillations are easily predicted with the help of time-dependent perturbation theory, 
which recovers the result obtained in \cite{BH0} for the symmetric case.
The physical picture is the one in which 
all $M$ atoms are transferred together, {\rd {as a single cluster,}} from one well to another, during a short period of time $\delta t \sim U^{-1}$,
while the precise moment of the transfer remains indeterminate, in accordance with the uncertainty principle.
\newline
This is achieved through destructive interference between other scenarios.
Compared to  the observation time, typically of the order of the large Rabi period,
the rapid transfer appears to be almost instantaneous. 
Accordingly, an observation  made in either well will almost certainly find there either $n$, or $n+M$ atoms, with all other counts occurring only rarely. While it is difficult to probe directly the duration of the collective transfer, a less direct proof
of this transfer mechanism is available.
 While the behaviour of the group of $M$ atoms is broadly similar to that of a 
single particle,  the composite nature of the group can be probed  if  tunnelling is switched on suddenly.
If so, the immediate transfer of individual atoms, or smaller groups of atoms, will result in a 
rapid oscillatory pattern, superimposed on the much slower Rabi cycle.
An analysis of the transition amplitude in Appendix A, as well as numerical results shown in Fig.4, 
show that these oscillations disappear, if tunnelling is turned on slowly, compared to $\delta t$,
and no splitting of the transferred group of atoms occurs at the start of the evolution.
For example, the one-atom-in-each well probability $p(1)$, shown in Fig. 4, ceases to oscillate, and assumes a
constant value in the adiabatic limit. This justifies the approximation of the integral in Eq.(\ref{B2}), and of all other 
oscillatory integrals in Appendix A, by a sum of end point contributions, and leaves the transition amplitude dominated
by the apparently simultaneous transfer of all atoms.
We note that the spectrum of these rapid oscillations typically contains combinations of all internal frequencies of the 
system, and does not lend itself to a simple analysis, except in the case $M=N=2$, discussed in Sec. IV. 
\newline
{\rd{ Finally, we expect the present theory, which treats collective tunnelling beyond the single-cluster picture, 
to find applications in 
experimental studies of such subtle aspects of the phenomenon, as quantum fluctuations in the observed averages, or the system's response to 
 temporal variations of its parameters.}}
\begin{center}
\textbf{Acknowledgements}
\end{center}
%One of us (DS) is grateful to the anonymous Referee of \cite{DS2} for suggesting to apply the path analysis to the problem at hand.
Financial support  of
MINECO and the European Regional Development Fund FEDER, through the grant
FIS2015-67161-P (MINECO/FEDER,UE), the Basque Government through Grant No IT986-16.
is also acknowledged  (DS and M.P), and of FIS2016-80681-P (MP).
%%%%%%%%%%%%%%%%%%%%%%%%%%%%%%%%
\section{Appendix A. Oscillatory integrals and the transition amplitude}
To illustrate the use of oscillatory integrals, discussed in Sec. III, we first consider a
 transition between states $|1\ra$ and $|2\ra$, with the energies $E_1$ and $E_2$, 
caused by a time-dependent perturbation $\hat{W}(t)=W(t)[|2\ra \la 1|+h.c.]$. 
If the system is in $|1\ra$ at $t=0$, the amplitude for finding it in $|2\ra$ at $t=T$, 
to the leading order in $\hat{W}(t)$, is given by
\begin{eqnarray} \label{B1}
A=-i\int_0^T \exp[-iE_2(T-t)]W(t)\exp(-iE_1t)dt\q\n
\equiv 
 -iT\exp(-iE_2T)\int_0^1 W(\tau) \exp[iT \Delta \tau]W(t)]d\tau,
\end{eqnarray}
where $\tau\equiv t \Delta$. 
%The period of oscillations of the exponential is, obviously, 
%$\delta t \sim 1/\Delta =1/|E_2-E_1|$.
 If the observation time 
is large, $T\Delta >> 1$, and $W(t)$ varies slowly, compared to $\delta t\sim 1/\Delta =1/|E_2-E_1|$, 
%but not compared to $T$, 
the main contributions to the integral
comes from the incomplete oscillations at the endpoints, 
i.e., from the regions  around $0$ and $T$, of a width $\sim \delta t$.
\begin{eqnarray} \label{B2}
A\approx -\Delta^{-1}[W(T) \exp(-E_1T)W(0)\exp(-iE_2T)].\q\q
\end{eqnarray}
%We note that Eq.(\ref{B2}) 
Now the amplitude $A$
has lost all the information about the overall 
behaviour of $W(t)$, and depends only on its initial value at $t=0$, 
and its current value at $t=T$. 
The short time it takes for the system to "jump" is, clearly $\sim \delta t$.
Moreover, if $W(t)$ is gradually switched
on from zero, the likelihood of finding the system in $|2\ra$, will 
depend only on the perturbation's current value $W(T)$. 
\subsection {Even number of atoms}
Next in complexity is the case in  which two atoms are transferred via a third state.
For simplicity, we consider a symmetric well, $\beta=0$, and start transferring an
even number of atoms $M=N=2K$, so that the minimum of the parabola in Fig. 2 is at an integer value of $n$.
 The transfer takes two steps, and the corresponding second-order amplitude
is given by 
\begin{eqnarray} \label{B1:2}
A\approx (-i)^2\exp(-iE_3T) \int_0^T dt_2 \int_0^{t_2}dt_1\times\q\q\n
\exp(i\Delta_2 t_2)W_2(t_2)
\exp(i\Delta_1 t_1)W_1(t_1),\q\q
\end{eqnarray}
where, as before, $\Delta_i=E_{i+1}-E_i$, the resonance condition reads $E_1=E_3$, 
and the "jump time" is $\delta t \sim 1/(E_2-E_1)$. The diagram in Fig.7 helps to visualise
the structure of the double integral in Eq.(\ref{B1:2}). Every integral, containing an exponential $\exp(i\Phi t)$, 
splits into contributions from its endpoints. An exception is the second integral in the left branch, 
where the phase $\Phi=\Delta_1+\Delta_2=0$ vanishes, and the integration has to be carried out 
over the whole interval $0\le t_2\le T$. This process, in which both atoms are transferred 
together at an unspecified moment $t_2$, contributes to the amplitude in Eq.(\ref{B1:2}) an amount
\begin{eqnarray} \label{B2:2}
A^{(I)}\approx i\Delta_1^{-1}\exp(-iE_3T) \int_0^T dt_2 W_2(t_2)
W_1(t_1).\q\q
\end{eqnarray}
There is also a process, in which the first atom jumps at $t=0$, and is followed by the second one at $t=T$
Its  contribution is
\begin{eqnarray} \label{B3:2}
A^{(II)}\approx -\Delta_1^{-2}\exp(-iE_2T) W_2(T)
W_1(0).\q\q
\end{eqnarray}
(Note that the phase is $E_2t$, since the system continues in the state $E_2$, 
until the second jump at $t=T$).
Finally, it is possible for both atoms to jump at $t=0$, and we have
\begin{eqnarray} \label{B4:2}
A^{(III)}\approx \Delta_1^{-2}\exp(-iE_3T) W_2(0)
W_1(0),\q\q
\end{eqnarray}
so that $A=A^{(I)}+A^{(II)}+A^{(III)}$. We note that if $W_1(t)$ is switched on 
from zero, slowly compared to the jump time  $\delta t$, the last two terms vanish, 
and the amplitude is fully determined by collective transfer of two atoms, 
$A=A^{(I)}$ at an unspecified time $t_2$.
The scheme in Fig. 7 is easily extended to collective transfer of more than two atoms, $M>2$
provided $N$ remains even. In general, such an amplitude is seen to be dominated by nearly 
instantaneous transfer of  $M$-atom cluster,
%the transition amplitude is evaluated in the $M$-th 
%order of the perturbation theory,
if  $W_i(t)$ varies slowly, compared to $\delta t=1/\text{min}\{\Delta_m\}$.
 With the observation times typically of order of the large Rabi period  $T_{Rabi}\sim W^{-M}$,
 the pictures should hold in most practical cases.
%%%%%%%%%%%%%%%%%%%%%%%%%
\begin{figure}[h]
\includegraphics[angle=0,width=8cm, height= 6cm]{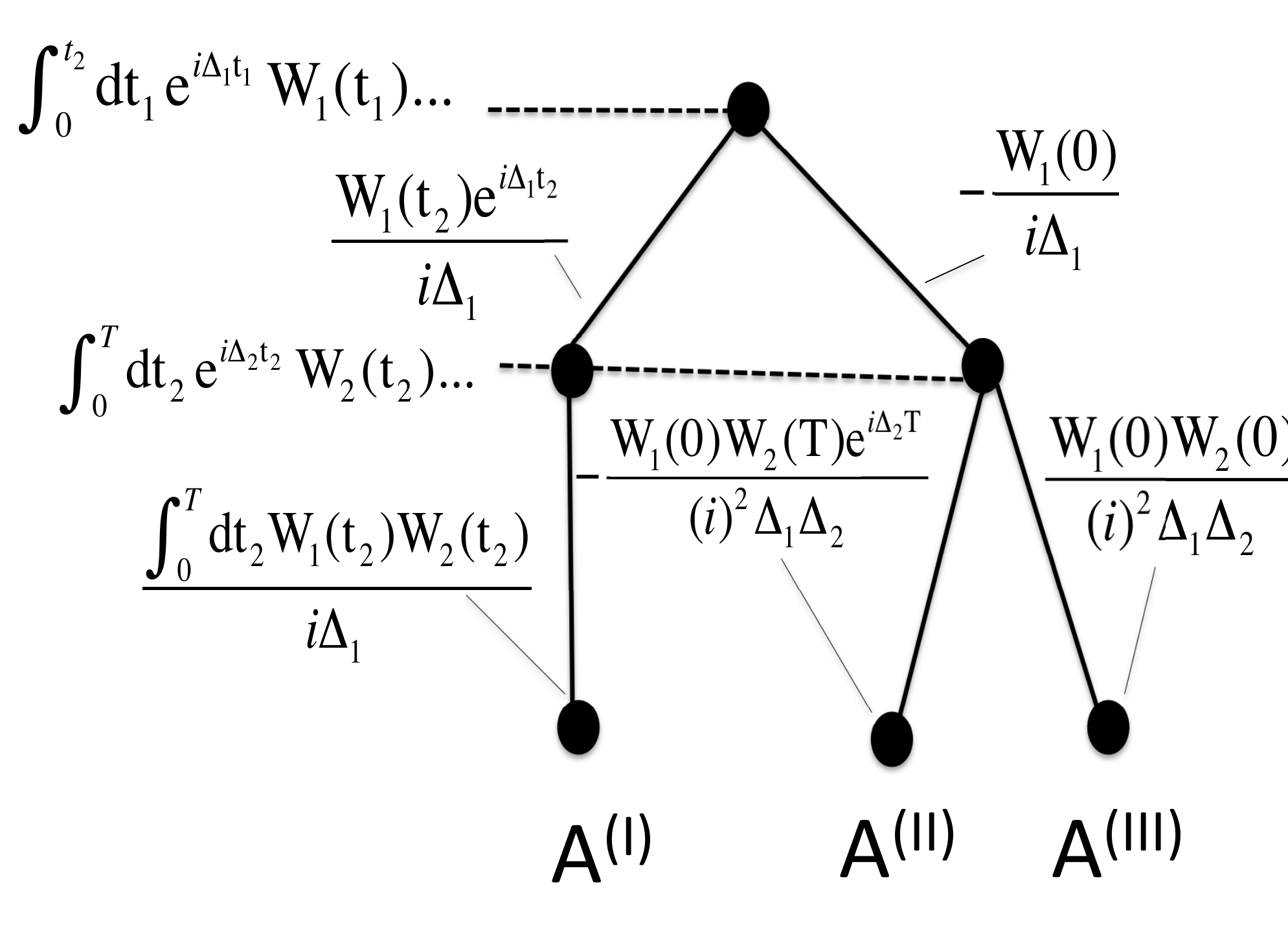}
\caption {A diagram illustrating evaluation of the double integral in Eq. (\ref{B1:2}).
% structure of the amplitude (\ref{B1:2}) for transmitting $M=2$ atoms.
Every integral, containing a rapidly oscillating exponential $\exp(i\Phi t)$,
is split into the contributions from its lower and upper limits. This does not, however, happen 
 in the last step of the left branch, where $\Phi=\Delta_1+\Delta_2=0$, and the integration is
 over $0\le t_3\le T$.}
\label{fig:FIG1}
\end{figure}
%%%%%%%%%%%%%%%%%%%%%%%%%%%%%%%%
%%%%%%%%%%%%%%%%%%%%%%%%%
\begin{figure}[h]
\includegraphics[angle=0,width=8cm, height= 6cm]{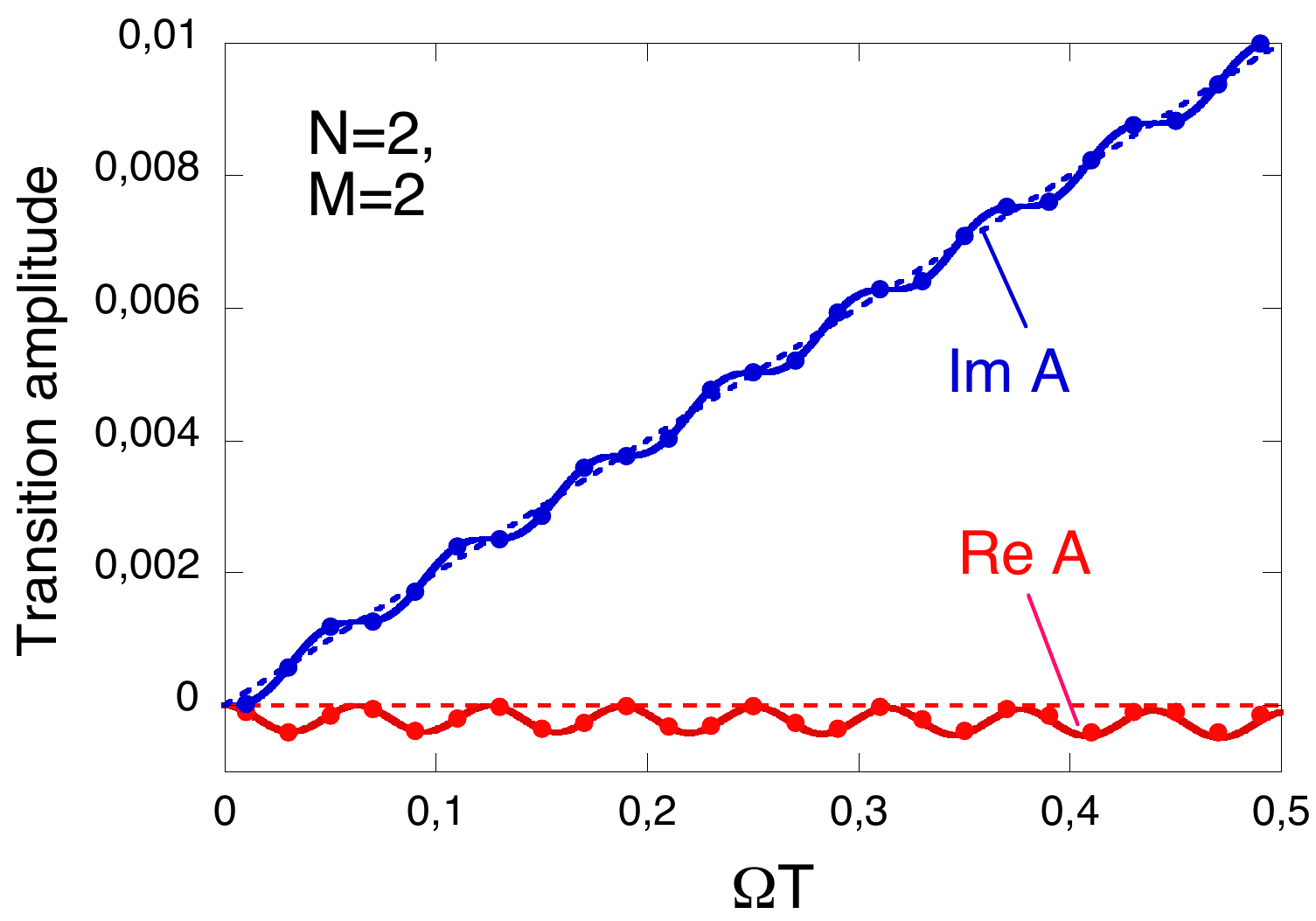}
\caption {Transmission of $M=2$ out of $N=2$ atoms.
{\rd {The exact (solid), and the second order (dots)}} amplitudes, given 
by Eqs.(\ref{A2}) and (\ref{B1:2}), respectively, coincide to graphical accuracy.
Also shown by dashed lines is $A^{(I)}$ in Eq.(\ref{B2:2}). 
As in Eqs.(\ref{A1a}), we use  $E_1=E_3=0$,
$E_2=-U$,constant values $W_1=W_2=\sqrt 2\Omega$, and $\Omega/U=0.01$.}
\label{fig:FIG1}
\end{figure}
%%%%%%%%%%%%%%%%%%%%%%%%%%%%%%%%
\subsection {Odd number of atoms}
%\newline
Although Eq.(\ref{0.4}) is valid for any $N$, 
the situation for $M=2K+1$, $K=1,2,...$, is slightly different, since now the 
two lowest states in Fig. 2, through which the system has to pass, are also in resonance. Thus, we have an 
integral
\begin{eqnarray} \label{B4}
A= (-i)^3\exp(-iE_4T) \int_0^{T}dt_3 \int_0^{t_3}dt_2 \int_0^{t_2}dt_1\times \q\q \n
\exp(i\Delta_3 t_3)W_3(t_3)\exp(i\Delta_2 t_2)W_2(t_2)\exp(i\Delta_1 t_1)W_1(t_1),
\end{eqnarray}
where $\Delta_2=0$, and $\Delta_1+\Delta_3=0$.
Four possible processes, which contribute to the amplitude (\ref{B4}) are shown in Fig.9.
The left branch on the diagram in Fig.7 corresponds, as before, to all atoms jumping at the 
same, yet unspecified, time, for which we have, 
\begin{eqnarray} \label{B5}
A^{(I)}=-i\exp(-iE_4T)\Delta_1^{-2}\times \n
\int_0^T W_1(t_3)W_2(t_3)W_3(t_3)dt_3.
\end{eqnarray}
The process in which the first two atoms jump at $t=0$, and the third at $t=T$, 
contributes 
\begin{eqnarray} \label{B6}
A^{(II)}=i\exp(-iE_3T)\Delta_1^{-2}\Delta_3^{-1}\times \n
%\times \q\q\q\n
 W_1(0)W_2(0)W_3(T), 
\end{eqnarray}
while all three atoms jumping at $t=0$ add
\begin{eqnarray} \label{B7}
A^{(III)}=-i\exp(-iE_4T)\Delta_1^{-2}\Delta_3^{-1}\times \n
%\times \q\q\q\n
 W_1(0)W_2(0)W_3(0). 
\end{eqnarray}
Finally, it is possible for the first atom to jump at $t=0$, 
for the second one to jump at some $0\le t_2\le T$, 
and for the third one to complete the transition at $t=T$.
The amplitude for this is
\begin{eqnarray} \label{B8}
A^{(IV)}=-i\exp(-iE_4T)\Delta_1^{-1}\Delta_3^{-1}
\times \n
 W_1(0)W_3(T)\exp(i\Delta_3T)\int_0^TW_2(t_2)dt_2. 
\end{eqnarray}
The amplitude for the fifth process in Fig.9 vanishes, since it involves
$\int_0^0W(t_2)dt_2=0$, and the full amplitude is the sum of four terms
$A=A^{(I)}+A^{(II)}+A^{(III)}+A^{(IV)}$.
Typically, all $W_i$ are of the same order of magnitude, $W_1\sim W_2\sim W_3\sim W$, 
and in the strong interaction limit, $W\delta t << 1$, the second and the third terms, 
which contain an extra factor of $\Delta_3^{-1}$, can be omitted. 
For all $W_i$ chosen to be constant, both $A^{(I)}$ and $A^{(IV)}$ grow linearly with time, 
\begin{eqnarray} \label{B9}
 A^{(I)} \sim T, \q A^{(IV)}\sim T \exp(i\Delta_3T), 
\end{eqnarray}
as shown in the inset of Fig.10.
However, $A^{(IV)}$ rapidly oscillates, compared to the large Rabi period, and can be omitted, 
when the Rabi frequency is calculated from the rate of change of $A$. 
Although not immediately seen from Eq.(\ref{B8}), this becomes clear from the graphs in
the main panel of Fig. 10. The oscillations do not increase proportional to time, and the linear
growth of the amplitude is determined, as one would expect, only by collective transfer of all 
three atoms. As before, switching tunnelling on slowly from zero, helps avoiding the splitting of the 
tunnelling cluster, and eliminates $A^{(II)}$, $A^{(III)}$, and $A^{(IV)}$, all proportional to $W_1(0)$.
The diagram in Fig.9 can be extended to the transfer of more than three atoms, by adding integrations and branches, as appropriate. In general, the amplitude would contain the leading term, describing nearly simultaneous 
transfer of all $M$ atoms,  as well as the additional sub-amplitudes, if the cluster is split at $t=0$  by sudden onset 
of tunnelling.
%%%%%%%%%%%%%%%%%%%%%%%%%
\begin{figure}[h]
\includegraphics[angle=0,width=9cm, height= 7cm]{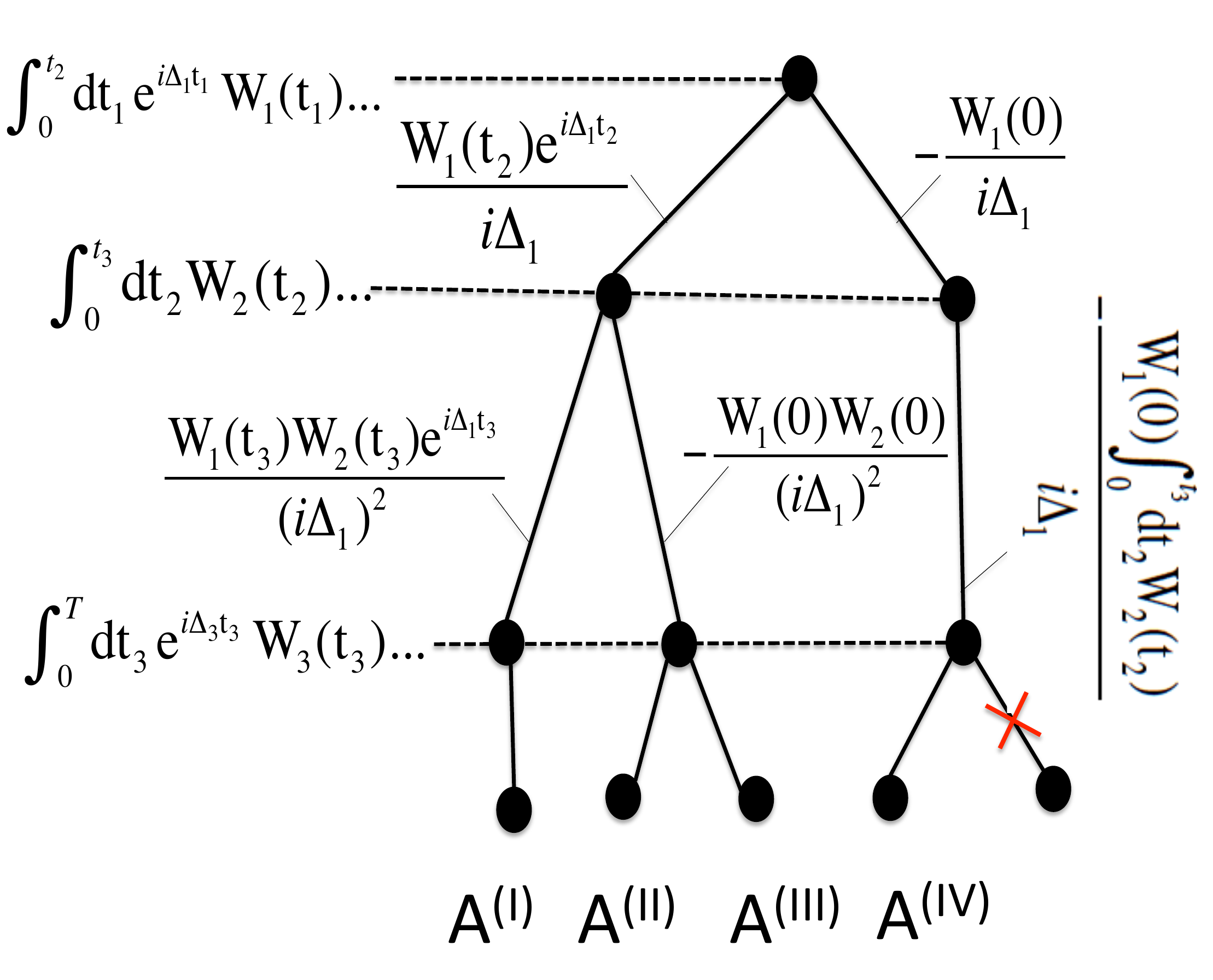}
\caption {Same as Fig. 7, except for the triple integral in Eq.(\ref{B4}).
The phase of the integrand vanishes in the last step of the left branch, 
due to the resonance condition $\Delta_1+\Delta_3=0$, and in the 
second step of the right branch, since $\Delta_2=0$.
In the right branch, the contribution from $t_3=0$ vanishes, as 
 marked by a cross.}
\label{fig:FIG1}
\end{figure}
%%%%%%%%%%%%%%%%%%%%%%%%%%%%%%%%
\begin{figure}[h]
\includegraphics[angle=0,width=9cm, height= 7cm]{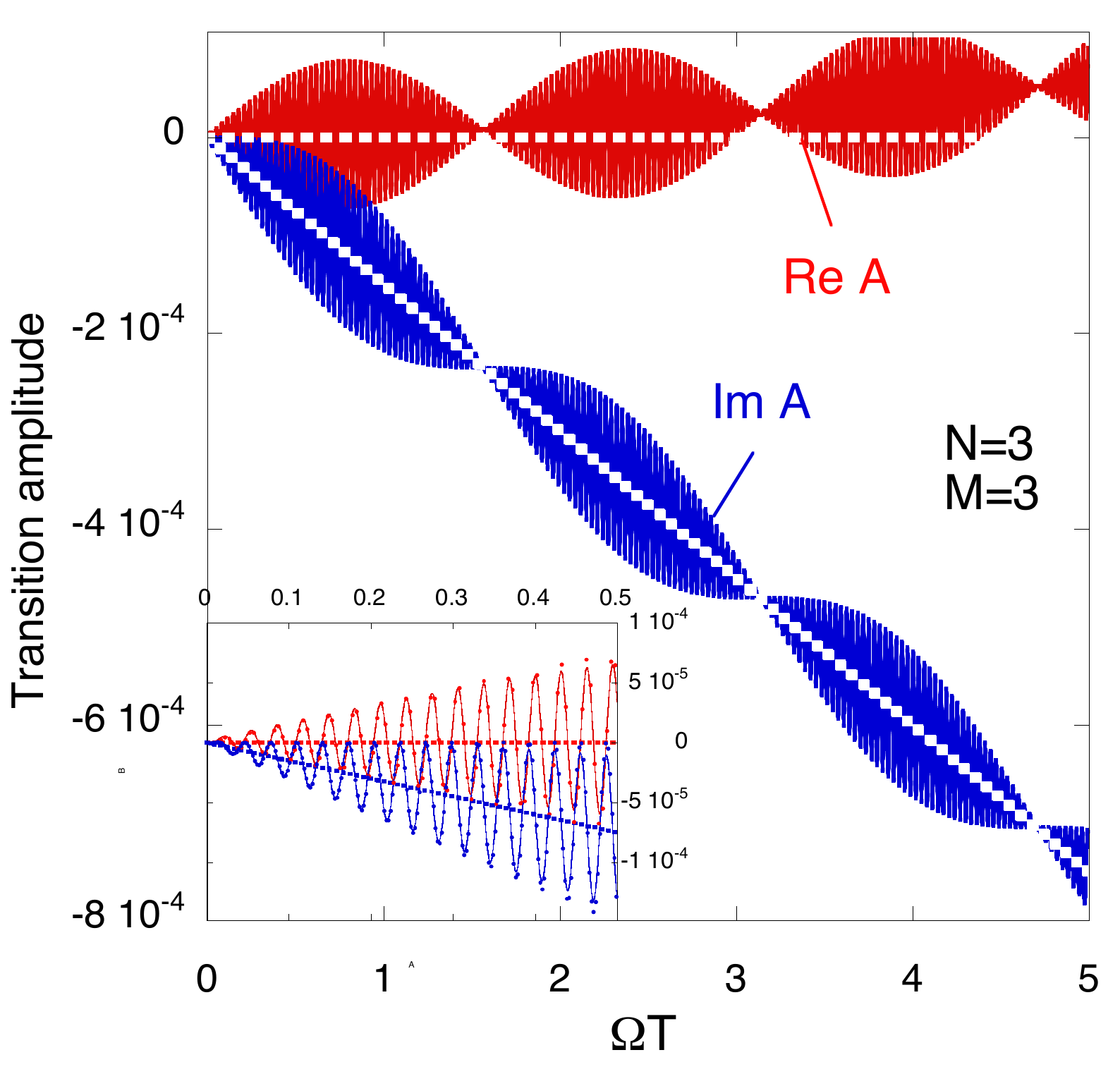}
\caption {Transmission of $M=3$ out of $N=3$ atoms.
Shown are the exact amplitude in Eq.(\ref{3.3}) (solid)  and $A^{(I)}$ in Eq.(\ref{B5}) (dashed). 
In the inset, he exact (solid) , and the second order amplitude in Eq.(\ref{B4}) (dashed)
are shown to coincide to graphical accuracy.
As in Eqs.(\ref{3.1}), we use  $E_1=E_4=0$,
$E_2=E_3=-2U$, constant values $W_1=W_3=\sqrt 3\Omega$, and $W_2=2\Omega$, and $\Omega/U=0.01$.}
\label{fig:FIG1}
\end{figure}
%%%%%%%%%%%%%%%%%%%%%%%%%%%%%%%%
\section{Appendix B. Exactly solvable cases} 
To write down a formal analytic solution to Eqs.(\ref{1.2}),
 we need to diagonalise the Hamiltonian matrix
 $H_{mn}\equiv\la m|\hat{H}|n\ra$, $m,n=0,1,...,N$. 
 Since algebraic equations have analytical solutions up to the 
 fourth order, this can be done for no more than $N=3$ atoms,
 in the case of a general asymmetric well.
 \newline
 For a symmetric well, $\beta =0$, a solution can be found for up to $N=6$ 
 atoms. Additional symmetry of $H_{mn}$,
$H_{mn}=H_{N-m,N-n}$
requires that its eigenvectors $\underline{a}^{n}\equiv
(a_0^n,a_1^n,...,a_N^n)$
have a definite parity,
\begin{eqnarray} \label{A1}
a^n_m=\alpha a^n_{N-m},\quad \alpha = \pm 1. 
\end{eqnarray}
This, in turn, allows us to reduce the size of the matrix 
we need to diagonalise in order to obtain the eigenvalues of $H_{mn}$,
and an analytical solution can, in principle, be obtained for $N\le 7$.
Below we analyse the $N=2$ and $N=3$ cases, assuming $\beta = 0$.
\subsection{The two-particle case ($N=2$)}
For $N=2$, we write $A_{n,N-n}(t) =\exp(-iUt)b_n(t)$, 
%measuring the energy from $\Ep(0)=\Ep(2)=U$,  we need to solve
where $b_n(t)$ satisfy
\begin{eqnarray}\label{A1a}
i\partial_t b_0=\sqrt{2}\Omega b_1\\
\nonumber
i\partial_tb_1=\sqrt{2}\Omega b_0-Ub_1+\sqrt{2}\Omega b_2\\
\nonumber
i\partial_t b_2=\sqrt{2}\Omega b_1
\end{eqnarray}
with the initial condition
$b_i(t=0)=\delta_{i0}, \quad i=0,1,2.$
This is equivalent to a second order equation for $b_1$
\begin{eqnarray}\label{A1aa}
\partial_{tt}b_1-iU\partial_t b_1+4\Omega^2b_1=0
\end{eqnarray}
with the initial condition
$b_1(0)=0, \quad \partial_t b_1(0)=-\sqrt{2}i\Omega.$
The corresponding solution is
%$$b_1(t)=\frac{\sqrt{2}\Omega}{\sqrt{U^2/4+4\Omega^2}
%\exp(-iUt/2)\sin(\sqrt{U^2/4+4\Omega^2)
\begin{eqnarray}\label{A1b}
b_1(t)=\frac{-\sqrt{2}\Omega}{\epsilon_1-\epsilon_2}
[\exp(i\epsilon_1t)-\exp(i\epsilon_2t)]
\end{eqnarray}
where.
$ \epsilon_{1,2}=U/2\pm\sqrt{U^2/4+4\Omega^2}.$
For two other coefficients we then have
\begin{eqnarray} \label{A2}
b_2(t)=-\sqrt{2}i\Omega\int_0^t b_1(s)ds = 
\frac{2\Omega^2}{(\epsilon_1-\epsilon_2)}\times\n
[\exp(i\epsilon_1t)/\epsilon_1-\exp(i\epsilon_2t)/\epsilon_2]
+\frac{2\Omega^2}{\epsilon_1\epsilon_2}
\end{eqnarray}
and
$$b_0(t)=1+b_2(t).$$
Returning to the strong interaction case, $\Omega/U <<1$, we note then that
\begin{eqnarray} \label{A3}
\epsilon_1-\epsilon_2\approx \epsilon_1\approx U,
\q
\epsilon_2 \approx- 4\Omega^2/U,\q
\epsilon_1\epsilon_2\approx -4\Omega^2.\q
\end{eqnarray}
With this we have
\begin{eqnarray}\label{A4}
b_0=\exp(-i\omega_0t)\cos(\omega_0 t), \q\n
 b_2=-i\exp(-i\omega_0t)\sin(\omega_0 t),\n
\omega_0=2\Omega^2/U,\q\q\q\q\q\q\q
\end{eqnarray}
which agrees with Eq.(\ref{0.10}).
\newline
For the condensate prepared initially with one boson in each well
we must change the initial condition for $b_1(0)$ to
$b_1(0)=1, \quad \partial_{\tau} b_1(0)=iU.$
Hence we have
$$b_1(t)=\frac{U-\epsilon_2}{\epsilon_1-\epsilon_2}\exp(i\epsilon_1t)
+\frac{\epsilon_1-U}{\epsilon_1-\epsilon_2}\exp(i\epsilon_2t)$$
and
$b_0(t)=b_2(t)=-i\sqrt{2}\Omega\int_0^tb_1(s)ds.$
It is readily seen that in the strong interaction limit
$\Omega/U<<1$, when the energy separation from the two states 
corresponding to two bosons in the same  well grows as $U$, the 
system stays in its ground state $|1\ra$. Indeed, rapid oscillations
of  $b_1(t)\approx \exp(-iUt)$, guarantee that both 
$ b_0(t)$ and $b_2(t)$ remain of order of  $\Omega/U)$.
%%%%%%%%%%%%%%%%%%%%
\subsection{The three-particle case ($N=3$)}
For three bosons, $N=3$, in a symmetric well, $\beta=0$, 
to evaluate  $b_n=\exp(2iUt) A_{n,N-n}$, we need to solve 
\begin{eqnarray}\label{3.1}
i\partial_t b_0=\sqrt{3}\Omega b_1\\
\nonumber
i\partial_tb_1=\sqrt{3}\Omega b_0-2Ub_1+2\Omega b_2\\
\nonumber
i\partial_tb_2=2\Omega b_1-2Ub_2+\sqrt{3}\Omega b_3\\
\nonumber
i\partial_t b_3=\sqrt{3}\Omega b_2.
\end{eqnarray}
This is equivalent to 
\begin{eqnarray}\label{3.2}
i\partial_t b_0=\sqrt{3}\Omega b_1\\
\nonumber
\partial_{tt}b^{\pm}+2iU^\pm\partial_tb^{\pm}+3\Omega^2b^{\pm}=0\\
%\partial_{tt}b^{\pm}+2i[\pm \Omega -U]\partial_tb^{\pm}+3\Omega^2b^{\pm}=0\\
%\nonumber
%i\partial_{tt}b^-+2i[U-\Omega]\partial_tb^-+3\Omega^2b^-=0
\nonumber
i\partial_t b_3=\sqrt{3}\Omega b_2.
\end{eqnarray} 
where
\begin{eqnarray}\label{3.2a}
b^{\pm}\equiv b_1\pm b_2 \quad \text{and}\quad U^\pm\equiv \pm \Omega -U
\end{eqnarray} 
which are solved with appropriate initial conditions discussed below.
\newline
{\it Initially, no particles in the right well} ($n=0$): $b_n(0)=\delta_{n0},\quad n=0,1,..3.$
Introducing (subscripts $1$ and $2$ correspond to the 
$+$ and $-$ signs of the square root) 
\begin{eqnarray}\label{3.3a}
\omga^{\pm}_{1,2}=[U^\pm ]\pm\sqrt{({U^\pm})^2+3\Omega^2}
%\omga^{\pm}_{1,2}=[U\mp\Omega]\pm\sqrt{(U\mp\Omega)^2+3\Omega^2}
\end{eqnarray}
we have
\begin{eqnarray}\label{3.3}\nonumber
b_0=
%1+b_3\q\q\q\q\q\q\q\q\q\q\q\q\q\q\q\q\q\q\q\q\q\q\q\q\n
-\frac{3}{2}\Omega^2\bigg \{
\frac{\exp(i\omga_1^+t)}{\omga_1^+(\omga_1^+-\omga_2^+)}
-\frac{\exp(i\omga_2^+t)}{\omga_2^+(\omga_1^+-\omga_2^+)}\q\q\q\q\q\q\q\q\q\n
+\frac{\exp(i\omga_1^-t)}{\omga_1^-(\omga_1^--\omga_2^-)}
-\frac{\exp(i\omga_2^-t)}{\omga_2^-(\omga_1^--\omga_2^-)}\bigg\}\q\q\q\q\q\q\q\q\q\n
b_1=-\frac{\sqrt{3}}{2}\Omega\bigg\{\frac{\exp(i\omga_1^+t)-\exp(i\omga_2^+t)}{\omga_1^+-
\omga_2^+}+
\frac{\exp(i\omga_1^-t)-\exp(i\omga_2^-t)}{\omga_1^--
\omga_2^-}\bigg\}\q\q\n
b_2=-\frac{\sqrt{3}}{2}\Omega\bigg\{\frac{\exp(i\omga_1^+t)-\exp(i\omga_2^+t)}{\omga_1^+-
\omga_2^+}-
\frac{\exp(i\omga_1^-t)-\exp(i\omga_2^-t)}{\omga_1^--
\omga_2^-}\bigg\}\q\q\n
b_3=-\frac{3}{2}\Omega^2\bigg\{
\frac{\exp(i\omga_1^+t)}{\omga_1^+(\omga_1^+-\omga_2^+)}
-\frac{\exp(i\omga_2^+t)}{\omga_2^+(\omga_1^+-\omga_2^+)}\q\q\q\q\q\q\q\q\q\n 
%\q\q\q\q\q\q\n
-\frac{\exp(i\omga_1^-t)}{\omga_1^-(\omga_1^--\omga_2^-)}
%\q\q\q\q\q\\
+\frac{\exp(i\omga_2^-t)}{\omga_2^-(\omga_1^--\omga_2^-)}\bigg\}\q\q\q\q\q\q\q\q\q
%\q\q\q
\end{eqnarray} 
In the strong interaction limit
$\Omega/U<<1$ we have
\begin{eqnarray}\label{3.4}
\omga_1^{\pm}-\omga_2^{\pm}\approx \omga^\pm_{1}\approx 2U, \q
%\n
e^\pm_{2}\approx -3\Omega^2/2U\mp 3\Omega^3/2U^2\q
%,\n
%^{\pm}\equiv 2(U\pm\Omega)$$
%\omga_1^{\pm}\approx 3\Omega^2/2U^{\pm}-3\Omega^4/4(U^{\pm})^3\n
%\omga_1^{\pm}(\omga_1^{\pm}-\omga_2^{\pm})\approx 3\Omega^2
\end{eqnarray}
so that
\begin{eqnarray}\label{3.5}
b_0(t) \approx -\exp(-3i\Omega^2t/2U) \cos(3\Omega^3t/2U^2), \quad\q \n
b_3(t) \approx i\exp(-3i\Omega^2t/2U) \sin(3\Omega^3t/2U^2), \quad\q \n
b_1(t)\approx b_2(t) =
o(\Omega/U)\q\q\q\q\q\q\q\q\q
\end{eqnarray}
in accordance with $\omega_0 \approx 3\Omega^3/2U^2$,  predicted by Eq.(\ref{0.10}).
%%%%%%%%%%%%%%%%%%%%%%%%%
\newline
{\it Initially, one particle in the right well ($n=1$):}
$$b_n(0)=\delta_{n1},\quad n=0,1,..3.$$
%Introducing (subscripts $1$ and $2$ corespond to the
%$+$ and $-$ signs of the square root)
%$$\omga^{\pm}_{1,2}=-[U\pm\Omega]\pm\sqrt{(U\pm\Omega)^2+3\Omega^2}$$
Solving Eqs.(\ref{3.1}) with this initial condition we have
\begin{eqnarray}\label{3.6}
%\nonumber
b_0=-\frac{\sqrt 3\Omega}{2}\times \q\q\q\q\q\q\q\q\q\q\n
%-i2^{-1}\n
\bigg\{-\frac{\omga_2^++2U^+}{\omga_1^+(\omga_1^+-
\omga_2^+)}\exp(i\omga_1^+t)+ 
\frac{\omga_1^++2U^+}{\omga_2^+(\omga_1^+-
\omga_2^+)}\exp(i\omga_2^+t)\q\q\q\n
-\frac{\omga_2^-+2U^-}{\omga_1^-(\omga_1^--
\omga_2^-)}\exp(i\omga_1^-t)+
\frac{\omga_1^-+2U^-}{\omga_2^-(\omga_1^--
\omga_2^-)}\exp(i\omga_2^-t)\bigg\},\q\q\q\n
b_1=-\frac{1}{2}
%2^{-1}
\bigg\{-\frac{\omga_2^++2U^+}{\omga_1^+-
\omga_2^+}\exp(i\omga_1^+t)+
\frac{\omga_1^++2U^+}{\omga_1^+-
\omga_2^+}\exp(i\omga_2^+t)\q\q\q\n
-\frac{\omga_2^-+2U^-}{\omga_1^--
\omga_2^-}\exp(i\omga_1^-t)+
\frac{\omga_1^-+2U^-}{\omga_1^--
\omga_2^-}\exp(i\omga_2^-t)\bigg\}\q\q\q\n
\end{eqnarray}
\begin{eqnarray}\label{3.6}\nonumber
b_2=-\frac{1}{2}
%2^{-1}
\bigg\{-\frac{\omga_2^++2U^+}{\omga_1^+-
\omga_2^+}\exp(i\omga_1^+t)+
\frac{\omga_1^++2U^+}{\omga_1^+-
\omga_2^+}\exp(i\omga_2^+t)\q\q\q\n
+\frac{\omga_2^-+2U^-}{\omga_1^--
\omga_2^-}\exp(i\omga_1^-t)-
\frac{\omga_1^-+2U^-}{\omga_1^--
\omga_2^-}\exp(i\omga_2^-t)\bigg\}\q\q\q\n
b_3=-\frac{\sqrt 3\Omega}{2}\times \q\q\q\q\q\q\q\q\q\q\n
%-i2^{-1}\n
\bigg\{-\frac{\omga_2^++2U^+}{\omga_1^+(\omga_1^+-
\omga_2^+)}\exp(i\omga_1^+t)+
\frac{\omga_1^++2U^+}{\omga_2^+(\omga_1^+-
\omga_2^+)}\exp(i\omga_2^+t)\q\q\q\n
+\frac{\omga_2^-+2U^-}{\omga_1^-(\omga_1^--
\omga_2^-)}\exp(i\omga_1^-t)-
\frac{\omga_1^-+2U^-}{\omga_2^-(\omga_1^--
\omga_2^-)}\exp(i\omga_2^-t)\bigg\}.\q\q\q
\end{eqnarray}
%where $U^\pm= U\pm \Omega$.
(Note that the expressions for $b_0$ and $b_3$ do not contain the 
time-independent terms obtained in integrating $b_1$ and $b_2$.
Since the hamiltonian matrix has four non-zero
eigenvalues, $b_0$ and $b_3$ cannot contain an additional
zero frequency.)
In the strong interaction limit
$\Omega/U<<1$ we have
%$$\omga_2^{\pm}\approx -2(V\pm\Omega) $$
%so that
\begin{eqnarray}\label{3.7}
b_1(t) \approx -\exp(2iUt)\cos(2\Omega t), \n
b_2(t) \approx -i\exp(2iUt)\sin(2\Omega t), \n
 b_0(t)\approx b_3(t) =O(\Omega/U),\q\q
 \end{eqnarray}
in accordance with $\omega_1 \approx 2\Omega$, obtained from Eq.(\ref{0.10}).


\begin{thebibliography}{10}

 \bibitem{Rabi}  I. I. Rabi, S. Millman, P. Kusch, J. R. Zacharias, {\it Phys. Rev.} {\bf 1939}, \textit{55}, 526.
   %%%%%%%%%%%%%%%%%%%%%%%%%%%%%%%%%%%%%%%
 \bibitem{Dud}  Y. O. Dudin, L. Li, F. Bariani, A. Kuzmich, {\it Nature Physics} {\bf 2012}, {\it 8}, 790.
  %%%%%%%%%%%%%%%%%%%%%%%%%%%%%%%%%%%%%%%
\bibitem{Bloch} S. Foelling, S. Trotzky, P. Cheinet, M. Feld, R. Saers, A. Widera, T. Mueller, I. Bloch,
{\it Nature} {\bf 2007}, {\it 448}, 1029.
 %%%%%%%%%%%%%%%%%%%%%%%%%%%%%%%%%%%%%%%

 %%%%%%%%%%%%%%UFAST%%%%%%%%%%%%%%%%%%%%
 \bibitem{Raiz} M. G. Raizen, S. Wan, C. Zhang, Q. Niu. {\it Phys. Rev. A} {\bf 2009}, {\it 80}, 030302(R).
 %%%%%%%%%%%%%%%
 \bibitem{MB1} 
K. Sakmann, A. I. Streltsov, O. E. Alon, L. S. Cederbaum, {\it  Phys. Rev. Lett.} {\bf 2009}, {\it  103}, 220601.
 %%%%%%%%%%%%%%%
 \bibitem{MB2} 
R. l. Beinke, S. Klaiman, L. S. Cederbaum, A. I. Streltsov,1,  O. E. Alon,  {\it  Phys. Rev. A}  {\bf 2015}, {\it  92}, 043627.
 %%%%%%%%%%%%%%%%
 \bibitem{MB3} 
J. Erdmann, S.I. Mistadikis, P. Schmelcher,  {\it  Phys. Rev. A} {\bf 2018},  {\it  98}, 053614.
 %%%%%%%%%%%%%%%
 \bibitem{BH1} R. Ma, M. E. Tai, P. M. Preiss, W. S. Bakr, J. Simon,  M. Greiner, {\it  Phys. Rev. Lett.} {\bf 2011}, {\it  107}, 095301.
  %%%%%%%%%%%%%%%
 \bibitem{BH2} F. Meinert,  M. J. Mark, E. Kirilov, K. Lauber, P. Weinmann, M. Groebner, A. J. Daley, H. C. Naegerl, {\it  Science} {\bf 2014}, {\it  344}, 1259
{\rd{ \bibitem{Ref2_5} S. I. Mistakidis, L. Cao, P. Schmelcher, {\it Mol. and Opt. Phys.} {\bf 2014}, {\it 47}, 225303. }}
 %%%%%%%%%%%%%%%%%%%%%%%%%%%%%%%%%%%%%%%
{\rd{ \bibitem{Ref2_6} S. I. Mistakidis, L. Cao, P. Schmelcher, {\it Phys. Rev, A} {\bf 2015} {\it 91}, 033611. }}
 %%%%%%%%%%%%%%UFAST%%%%%%%%%%%%%%%%%%%%
{\rd{\bibitem{Ref2_7} J. Neuhaus-Steinmetz, S. I. Mistakidis, P. Schmelcher, {\it Phys. Rev. A} {\bf 2017}, {\it 95}, 053610. }}
%\bibitem{Legg} A.J. Leggettt,  Rev. Mod. Phys., . {\bf 73}, 307 (2001). 
 {\rd{\bibitem{Ref1_1} L. Cao, I. Brouzos, B. Chatterjee, P. Schmelcher, {\it New. J. Phys.} {\bf 2012}, {\it 14}, 093011.}}
  %%%%%%%%%%%%%%%%%%%%%%%%%%%%%%%%%%%%%%%
 {\rd{\bibitem{Ref1_2} R. E. Barfknecht, A. Foerster, N. T. Zinner, {\it New. J. Phys.} {\bf 2018}, {\it 20}, 063014.}}
  %%%%%%%%%%%%%%%%%%%%%%%%%%%%%%%%%%%%%%
 {\rd{\bibitem{Ref2_2} G.J. Milburn, J. Corney, E.M. Wright, D. F. Walls, {\it Phys. Rev. A} {\bf 1997}, {\it 55}, 4318.}}
   %%%%%%%%%%%%%%%%%%%%%%%%%%%%%%%%%%%%%%%
{\rd{\bibitem{Ref2_3} A. Smerzi, S. Fantoni, S. Giovanazzi, S. R. Shenoy, {\it Phys. Rev. Lett.}, {\bf 1997},{\it 79}, 4950. }}
   %%%%%%%%%%%%%%%%%%%%%%%%%%%%%%%%%%%%%%%
{\rd{\bibitem{Ref2_1} M. Albiez, R. Gati, J. F\"olling, S. Hunsmann, M. Cristiani, M. K. Oberthaler, {\it Phys. Rev. Lett.}  {\bf 2005}, {\it 95}, 010402.}}
  %%%%%%%%%%%%%%%%%%%%%%%%%%%%%%%%%%%%%%%
{\rd{\bibitem{Ref2_4} Y. Chen, S. Nascimb\`ene, M. Aidelsburger, M. Atala, S. Trotzky, I. Bloch, {\it  Phys. Rev. Lett.}, {\bf 2011}, {\it107}, 210405.}}
   %%%%%%%%%%%%%%%%%%%%%%%%%%%%%%%%%%%%%%%
%\bibitem{Legg} A.J. Leggettt,  Rev. Mod. Phys., . {\bf 73}, 307 (2001).


%%%%%%%%%%%%%%%%
\bibitem{BH0} G. Kalosaka, A. R. Bishop,  V. M. Kenkre,  {\it  J. Phys. B.}, {\bf 2003}, {\it  36}, 3233.
%%%%%%%%%%%%%%%
\bibitem{Bern} L. Bernstein, J. C. Eilbeck, A. C. Scott, {\it  Nonlinearity}, {\bf 1990}, {\it  3}, 293.
%%%%%%%%%%%%%%%
\bibitem{BH00} G. Kalosaka, A. R. Bishop,  V. M. Kenkre,  {\it  Phys. Rev. A}, {\bf 2003}, {\it  68}, 023602.
%%%%%%%%%%%%%%%
\bibitem{FOOTall} 
Imagine there are two rooms, one of which is full of people. The people move from one room to the other, and each time one checks, all people are found in the same room. This  implies that whenever people decide to move, they do it not one by one, but all together. The same must be the case with the bosons.
%%%%%%%%%%%%%%%%%%%
\bibitem{Feyn} R. P. Feynman and A. R. Hibbs, {\it Quantum Mechanics and Path Integrals, McGraw-Hill, New York} {\bf 1965}. 
%%%%%%%%%%%%%%%%%%%
%%%%%%%%%%%%%%%%
%\bibitem{NONLIN} 
%L. Bernstein, J.C. Eilbeck, A.C. Scott, 
%Nonlinearity, {\bf 3}, 293 (1990).

\end{thebibliography}
\end{document}